\documentclass[a4paper,11pt]{article}
\pdfoutput=1
\usepackage{jheppub}
\usepackage{graphicx}
\usepackage{epsfig}
\usepackage{subfig}
\usepackage{url}
\usepackage{dsfont}
\usepackage{amssymb}
\usepackage{slashed}
\usepackage{booktabs,multirow,tabularx}
\usepackage{braket}
\usepackage[table]{xcolor}
\usepackage{tabularx}
\usepackage{array}
\usepackage{ulem}
\usepackage{comment}

\DeclareMathAlphabet{\pazocal}{OMS}{zplm}{m}{n}

\newcommand\ep{\epsilon}

\def\p0{{\bigl.^3\hspace{-1mm}P^{[8]}_0}}

\def\to{\rightarrow}

\def\bqa{\begin{eqnarray}}
\def\eqa{\end{eqnarray}}
\def\bc{\begin{center}}
\def\bc{\end{center}}

\newcommand\ResNOne{{\rm N}_1}
\newcommand\ResNTwo{{\rm N}_2}
\newcommand\ResNbOne{\overline{{\rm N}}_1}
\newcommand\ResNbTwo{\overline{{\rm N}}_2}
\newcommand\NkLONkLLtimesNLOmt{{\rm (N}^{k}{\rm LO+N}^k{\rm LL)\otimes NLO}_{m_t}}
\newcommand\NkLOtimesNLOmt{{\rm N}^{k}{\rm LO \otimes NLO}_{m_t}}
\newcommand\NkLONkLL{{\rm N}^{k}{\rm LO+N}^k{\rm LL}}
\newcommand\NkLO{{\rm N}^{k}{\rm LO}}
\newcommand\NLOmt{{\rm NLO}_{m_t}}
\newcommand\NNLOtimesNLOmt{{\rm NNLO\otimes NLO}_{m_t}}
\newcommand\NNLONNLLtimesNLOmt{{\rm (NNLO+NNLL)\otimes NLO}_{m_t}}
\newcommand\NtLOtimesNLOmt{{\rm N}^{3}{\rm LO \otimes NLO}_{m_t}}
\newcommand\NtLONtLLtimesNLOmt{{\rm (N}^{3}{\rm LO+N}^3{\rm LL)\otimes NLO}_{m_t}}

\def\be{\begin{equation}}
\def\ee{\end{equation}}
\def\bea{\begin{eqnarray}}
\def\eea{\end{eqnarray}}

\def\dfrac{\displaystyle\frac}

\newlength{\irrepwidth}
\newlength{\irrepbarthickness}
\newlength{\irrepbarheight}
\def\primes#1#2{\count0=#1 \loop \ifnum\count0>0 \advance\count0 by -1 #2\repeat}

\usepackage{tikz}
\usetikzlibrary{positioning,arrows}
\usetikzlibrary{decorations.pathmorphing}
\usetikzlibrary{decorations.markings}
\usetikzlibrary{shapes.geometric}
\usepackage{endnotes}
\tikzset{
    vector/.style={decorate, decoration={snake}, draw},
    provector/.style={decorate, decoration={snake,amplitude=2.5pt}, draw},
    antivector/.style={decorate, decoration={snake,amplitude=-2.5pt}, draw},
    fermion/.style={draw=black,
      postaction={decorate},decoration={markings,mark=at position .55
        with {\arrow[draw=black]{>}}}},
    fermionbar/.style={draw=black, postaction={decorate},
                       decoration={markings,mark=at position .55 with {\arrow[draw=black]{<}}}},
    fermionnoarrow/.style={draw=black},
    gluon/.style={decorate, draw=red,decoration={coil,amplitude=4pt, segment length=6pt}},
    scalar/.style={dashed,draw=black,
      postaction={decorate},decoration={markings,mark=at position .55
        with {\arrow[draw=black]{>}}}},
    scalarbar/.style={dashed,draw=black,
      postaction={decorate},decoration={markings,mark=at position .55
        with {\arrow[draw=black]{<}}}},
    scalarnoarrow/.style={dashed,draw=black},
    electron/.style={draw=black,
      postaction={decorate},decoration={markings,mark=at position .55
        with {\arrow[draw=black]{>}}}},
    bigvector/.style={decorate, decoration={snake,amplitude=4pt}, draw},
}

\normalbaselines

\title{N$^3$LO+N$^3$LL QCD improved Higgs pair cross sections}

\author[]{A.H. Ajjath}

\author[]{and Hua-Sheng Shao}

\affiliation[]{Laboratoire de Physique Th\'eorique et Hautes Energies (LPTHE), UMR 7589, Sorbonne Universit\'e et CNRS, 4 place Jussieu, 75252 Paris Cedex 05, France}

\emailAdd{aabdulhameed@lpthe.jussieu.fr}
\emailAdd{huasheng.shao@lpthe.jussieu.fr}

\abstract{We report a new calculation of the soft-gluon threshold resummation for the Higgs boson pair production in the dominant production mode -- gluon-gluon fusion -- up to the next-to-next-to-next-to-leading logarithmic (N$^3$LL) accuracy. After matching N$^3$LL to the next-to-next-to-next-to-leading order (N$^3$LO) QCD calculation in the infinite top quark mass approximation, we show that the central values of the inclusive cross sections are quite stable with respect to N$^3$LO, while the conventional renormalisation and factorisation scale uncertainties are reduced by a factor of two, reaching to the subpercent level. Our study further consolidates the good asymptotic perturbative convergence. After combining with the full top-quark mass dependent next-to-leading order QCD results, our most advanced predictions are presented for both the inclusive total cross sections and the differential invariant mass distributions of the Higgs pair.}


\begin{document}
\maketitle
\flushbottom
\section{Introduction}

The studies of the $125$ GeV Higgs boson $h$ at the Large Hadron Collider (LHC) turn into the second decade since its discovery~\cite{ATLAS:2012yve,CMS:2012qbp}. Many Higgs properties, such as its mass, width, spin and CP, have been thoroughly investigated. Its interactions with the force carriers ($Z,W,\gamma$, and gluon $g$) and the third-generation charged fermions ($t,b,\tau$) have all been established, which are compatible with the Standard Model (SM) predictions within $\sim 10\%$ accuracy~\cite{ATLAS:2022vkf,CMS:2022dwd}~\footnote{A caveat: the quoted uncertainties may vary depending on the assumptions that have been made.}. The observation of the Yukawa coupling between the Higgs boson and the first second-generation fermion muon is in reach. One of the central questions remaining to be answered at the LHC is that whether the Higgs boson interacts with itself. A remarkable prediction of the SM is that, analogous to the way of Higgs boson interacting with other massive elementary particles, the self-interaction strength of the Higgs boson scales with its mass. The  latter has been measured to be around $0.1$ to $0.2\%$ accuracy. The knowledge of the Higgs self interactions is essential to scrutinize the Higgs field potential, a key to understand the electroweak spontaneous symmetry breaking mechanism and on how the subatomic particles acquire their masses. The only thing we know about the potential so far is from the mass $m_h$ of the Higgs boson, which is the second derivative of the potential with respect to the Higgs field around the (local) minimum of potential, at which the Higgs field averagely takes the non-zero vacuum expectation value $v\approx 246.2$ GeV.

The shape of the Higgs potential has in fact the far reaching influences on cosmology (see e.g., refs.~\cite{Chen:2020pma,Salam:2022izo}). The distinct fates of our universe are determined by the vacuum stability or instability of the potential. In the former case, the current minimum at $v\approx 246.2$ GeV is the global minimum, while it is just a local minimum in the latter case. Thus, our universe will decay through quantum tunnelling at some point in the future if the vacuum is unstable~\cite{Isidori:2001bm}. This is the case predicted in the SM~\cite{Degrassi:2012ry,Alekhin:2012py,Branchina:2013jra,DiLuzio:2015iua}. Such an instability can be cured in many beyond the SM (BSM) theories. The role of the Higgs field is also crucial in the early universe. In particular, at around a nanosecond after the big bang, the universe underwent from the electroweak symmetry unbroken phase to the broken phase~\cite{Dine:1992wr}. Such a phase transition could be a first or second order as well as a crossover depending on the Higgs potential. Lattice calculations~\cite{Kajantie:1996mn,Csikor:1998eu} show that it should be a crossover in the setup of the SM. The (strong) first-order phase transition provides one necessary condition~\cite{Shaposhnikov:1987tw} for generating matter-antimatter asymmetry at the electroweak scale. Besides, the first-order transition may leave a few detectable cosmic imprints, such as primordial~\footnote{The word ``primordial" here means that the gravitational waves and magnetic fields are not produced by the astronomical objects. They are really leftovers from the cosmic evolution.} gravitational waves~\cite{Witten:1984rs,Hogan:1986qda} and magnetic fields~\cite{Grasso:2000wj}, and topological defects.

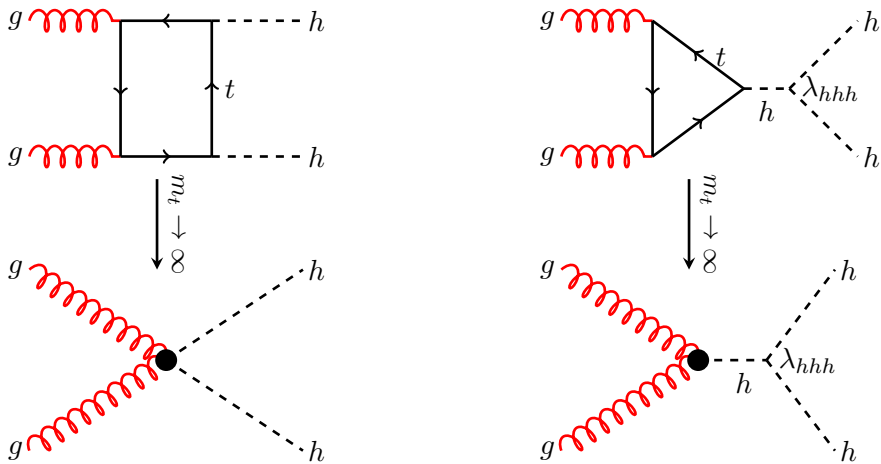
\begin{figure}[h!]
  \begin{tikzpicture}[line width=1 pt, scale=0.6]
  \hspace{1.5cm}
\draw[gluon] (-2,1.5) -- (0,1.5);
\draw[gluon] (-2,-1.5) -- (0,-1.5);
\draw[fermion] (0,1.5) -- (0,-1.5);
\draw[fermion] (0,-1.5) -- (2,-1.5);
\draw[fermion] (2,-1.5) -- (2,1.5);
\draw[fermion] (2,1.5) -- (0,1.5);
\draw[scalarnoarrow] (2,1.5) -- (4,1.5);
\draw[scalarnoarrow] (2,-1.5) -- (4,-1.5);
\node at (-2.3,1.5) {$g$};
\node at (-2.3,-1.5) {$g$};
\node at (4.3,1.5) {$h$};
\node at (4.3,-1.5) {$h$};
\node at (2.4,0) {$t$};
\vspace{2cm}
\draw [-stealth](0.8,-2) -- (0.8,-4);
\node[font = {\small},rotate=270,above] at (0.8,-3) {$m_t \rightarrow \infty$};
\draw[gluon] (-2,-4) -- (1,-6);
\draw[gluon] (-2,-8) -- (1,-6);
\filldraw [black] (1,-6) circle (6pt);
\draw[scalarnoarrow] (1,-6) -- (4,-4);
\draw[scalarnoarrow] (1,-6) -- (4,-8);
\node at (-2.3,-4) {$g$};
\node at (-2.3,-8) {$g$};
\node at (4.3,-4) {$h$};
\node at (4.3,-8) {$h$};
\hspace{7cm}
\draw[gluon] (-2,1.5) -- (0,1.5);
\draw[gluon] (-2,-1.5) -- (0,-1.5);
\draw[fermion] (0,1.5) -- (0,-1.5);
\draw[fermion] (0,-1.5) -- (2,0);
\draw[fermion] (2,0) -- (0,1.5);
\draw[scalarnoarrow] (2,0) -- (3,0);
\draw[scalarnoarrow] (3,0) -- (4.5,1.5);
\draw[scalarnoarrow] (3,0) -- (4.5,-1.5);
\node at (-2.3,1.5) {$g$};
\node at (-2.3,-1.5) {$g$};
\node at (4.8,1.5) {$h$};
\node at (2.5,-0.5) {$h$};
\node at (4.8,-1.5) {$h$};
\node at (1.5,0.7) {$t$};
\node at (3.9,0) {$\lambda_{hhh}$};
\vspace{2cm}
\draw [-stealth](0.8,-2) -- (0.8,-4);
\node[font = {\small},rotate=270,above] at (0.8,-3) {$m_t\to \infty$};
\draw[gluon] (-2,-4) -- (1,-6);
\draw[gluon] (-2,-8) -- (1,-6);
\filldraw [black] (1,-6) circle (6pt);
\draw[scalarnoarrow] (1,-6) -- (2.5,-6);
\draw[scalarnoarrow] (2.5,-6) -- (4,-4);
\draw[scalarnoarrow] (2.5,-6) -- (4,-8);
\node at (-2.3,-4) {$g$};
\node at (-2.3,-8) {$g$};
\node at (4.3,-4) {$h$};
\node at (2.0,-6.5) {$h$};
\node at (4.3,-8) {$h$};
\node at (3.4,-6) {$\lambda_{hhh}$};
\end{tikzpicture}
\caption{Representative lowest order Feynman diagrams of $gg\to hh$ in the SM with full top quark mass dependence (upper) and in the heavy/infinite top quark mass approximation (lower).} 
\label{fig:feynmandiag}
\end{figure}

As aforementioned, the crucial input to pin down the Higgs potential is by measuring the Higgs self interactions. These self interactions are accessible by studying the multiple-Higgs boson production from high-energy particle collisions, such as at the LHC~\cite{LHCHiggsCrossSectionWorkingGroup:2016ypw} or the future colliders~\cite{Cepeda:2019klc,Contino:2016spe}. The most viable production mode is the production of a pair of Higgs bosons. We will concentrate on the process of the Higgs boson pair production at hadron colliders in this paper. Such a process is vital in directly probing the Higgs trilinear self coupling $\lambda_{hhh}$~\cite{DiMicco:2019ngk} and in indirectly constraining the Higgs quartic self coupling~\cite{Borowka:2018pxx,Bizon:2018syu}, where in the SM $\lambda_{hhh}$ is uniquely fixed by the Higgs mass and the vacuum expectation value via $\lambda_{hhh}^{\rm SM}=\frac{m_h^2}{2v^2}\approx 0.129$. Due to the pattern of the couplings between the Higgs boson and other SM particles, Higgs boson is very hard to be experimentally detected (partially) because of its low production rates and the small branching fractions of the background-clean decay modes. The di-Higgs is more than doubling the challenges. The di-Higgs production cross sections are three orders of magnitude smaller than the single Higgs counterparts, in which the two Higgs bosons are predominantly produced in the fusions of two initial gluons emitted by protons (see figure~\ref{fig:feynmandiag}).

The paper will focus on quantifying the high precision Higgs pair production cross sections via the gluon-gluon fusion (ggF) channel. We want to stress that the target precision is strongly tied to determine how large portion of the BSM scenarios can be probed (cf., e.g., figure 1 in ref.~\cite{Huang:2016cjm}). A question of ``if we measure a deviation of the trilinear Higgs coupling with respect to the SM value $\delta_{\lambda}\equiv\kappa_{\lambda}-1\equiv \frac{\lambda_{hhh}}{\lambda_{hhh}^{\rm SM}}-1\neq 0$, what is the energy scale that the new physics lies ?" is pursued in refs.~\cite{Falkowski:2019tft,Chang:2019vez} . For instance, ref.~\cite{Chang:2019vez} finds the perturbation theory breaks at the scale $\lesssim \frac{13~{\rm TeV}}{|\delta_{\lambda}|}$ regardless of the shape of the Higgs potential. It means that a BSM dynamics should exist in the $130$ TeV ($1.3$ PeV) regime with a $10\%$ ($1\%$) deviation of $|\delta_{\lambda}|$ from $0$. Therefore, the capability of identifying a small BSM signature requires us to improve the theoretical systematic control in the SM.

The existing direct measurements of the Higgs boson pair cross sections at the LHC only loosely bound on the trilinear Higgs self-coupling $\lambda_{hhh}$ so far. The current best constraints $-1.24<\kappa_{\lambda}<6.49$ and $-0.6<\kappa_{\lambda}<6.6$ at $95\%$ confidence level were placed by the CMS~\cite{CMS:2022dwd} and ATLAS~\cite{ATLAS:2022jtk} collaborations respectively.\footnote{In ATLAS paper~\cite{ATLAS:2022jtk}, the best constraint by combining the Higgs pair cross section and the loop effects from the single Higgs process is  $-0.4<\kappa_{\lambda}<6.3$. The electroweak loop effects from the single Higgs production can indirectly probe $\lambda_{hhh}$. However at the HL-LHC, the indirect limit would be harder to be improved since it is dominated by the systematic errors.}
Although the results are very encouraging, the limit does not yet reach to most of the interesting regions of the vast BSM scenarios (see, e.g., refs.~\cite{Kanemura:2002vm,OConnell:2006rsp,Cacciapaglia:2017gzh,Jurciukonis:2018skr,Durieux:2022sgm}). It is even just touching the range $|\delta_{\lambda}|=|\kappa_{\lambda}-1|<5$ set by the perturbativity argument~\cite{Maltoni:2018ttu}. The precision of $\kappa_{\lambda}$ will be largely improved at the HL-LHC with 3000 fb$^{-1}$ integrated luminosity~\cite{Cepeda:2019klc} and the envisaged future hadron colliders~\cite{Contino:2016spe,deBlas:2019rxi,Mangano:2020sao}. $\delta_{\lambda}$ is estimated to be achievable better than $\pm 50\%$ (around $\pm 5\%$) at the HL-LHC (a future 100 TeV pp collider).~\footnote{There is, however, a caveat in order to interpret these numbers. The values of $\delta_{\lambda}$ were derived based on several assumptions extrapolated from what have been understood or used in the LHC Run 2 analysis, where the theoretical uncertainties have been assumed to be reduced by a factor of two with respect to the state of the art at the time.}

Before moving on, we briefly describe how the cross sections of the ggF process $gg\to hh$ are calculated in the SM. The perturbative calculations for the process are quite challenging due to the absence of the tree-level interactions between the Higgs boson and the gluon. The leading order (LO) cross section is derived from the square of the one-loop amplitude dominantly via the virtual top quark loop, where two representative Feynman diagrams can be found in the upper row of figure~\ref{fig:feynmandiag}. The higher-order QCD perturbative calculations are very difficult due to the complicated multi-loop multi-scale Feynman integrals. Thanks to the new technology development of the numerical approaches~\cite{Borowka:2015mxa,Li:2015foa,Dick2004493}, the next-to-leading order (NLO) results are presented in refs.~\cite{Borowka:2016ehy,Borowka:2016ypz,Baglio:2018lrj,Davies:2019dfy,Baglio:2020wgt,Baglio:2020ini}. The NLO calculation has been further improved either by soft-gluon resummation~\cite{Ferrera:2016prr,DeFlorian:2018eng} or after matching to parton showers~\cite{Heinrich:2017kxx,Jones:2017giv,Heinrich:2019bkc}. These are the best results achieved so far by taking into account the full top quark mass $m_t$ dependence. They leave a few theory challenges to be tackled in the future. First of all, the NLO results are plagued with the large theoretical uncertainties both from the scale variations~\cite{Borowka:2016ehy}
and the top-quark mass scheme dependence~\cite{Baglio:2018lrj}. Some differential
distributions may differ significantly by adopting different matching schemes~\cite{Heinrich:2017kxx} or using different shower scales~\cite{Jones:2017giv}.

A second useful way of improving the cross sections is applying the so-called infinite top quark mass approximation,~\footnote{It is also alternatively called the ``heavy top quark mass approximation" or ``large top quark mass approximation" or dubbed as the Higgs effective field theory (HEFT) approach in the literature.} denoting as $m_t\to \infty$. In such an approximation, the top quark loops shrink to the contact points (cf. the bullets in the lower row of figure~\ref{fig:feynmandiag}) by taking the heavy top quark mass approximation $m_t\gg m_h$ in the amplitude. Under this working approximation, we have circumvented the aforementioned complicated multi-loop Feynman integrals. One is therefore able to go beyond much further than the loop-induced case, albeit still with a lot of theoretical efforts needed. The state-of-the-art perturbative QCD calculation in the $m_t\to \infty$ approximation is next-to-next-to-next-to-leading order (N$^3$LO)~\cite{Chen:2019lzz,Chen:2019fhs}, which was certainly impossible without the decades of theoretical efforts~\cite{Dawson:1998py,deFlorian:2013uza,deFlorian:2013jea,Grigo:2014jma,deFlorian:2016uhr,Spira:2016zna,Gerlach:2018hen,Anastasiou:2015vya,Mistlberger:2018etf,Dulat:2018rbf,Catani:2007vq,Banerjee:2018lfq,Gehrmann:2012ze,Gehrmann:2014yya,Echevarria:2016scs,Luo:2019bmw,Luo:2019hmp,Li:2016ctv,Alwall:2014hca,Frederix:2018nkq} by the whole community. Although N$^3$LO QCD corrections only enhance the central values of next-to-next-to-leading order (NNLO) by $3\%$, the renormalisation and factorisation scale uncertainties are dramatically reduced by a factor of $3$ to $4$ to reach to the percent level accuracy. The soft-gluon resummation effects have been studied up to NNLO plus next-to-next-to-leading logarithmic (NNLL) accuracy in ref.~\cite{deFlorian:2015moa}.~\footnote{An earlier threshold resummation at a lower accuracy was carried out in ref.~\cite{Shao:2013bz}.} The finite top quark mass corrections can be included either by taking into account more terms in the $\frac{m_h^2}{m_t^2}$ expansion~\cite{Grigo:2013rya,Grigo:2015dia,Degrassi:2016vss,Davies:2019xzc,Davies:2021kex,Davies:2019dfy} or by rescaling $m_t\to\infty$ amplitudes (cross sections) with the known full-$m_t$ dependent amplitudes (cross sections)~\cite{Maltoni:2014eza,Frederix:2014hta,Grazzini:2018bsd,Chen:2019fhs}. Moreover, there are also many interesting attempts to compute the two-loop ggF di-Higgs amplitudes in the analytic forms by taking various approximations~\cite{Davies:2018ood,Davies:2018qvx,Bonciani:2018omm,Xu:2018eos,Bellafronte:2022jmo}. Recently, partial results of the NLO electroweak corrections, namely from the top Yukawa coupling induced, become available~\cite{Davies:2022ram,Muhlleitner:2022ijf}.

In this work, we further improve the perturbative QCD calculations of $gg\to hh$ with the threshold resummation up to next-to-next-to-next-to-leading logarithmic (N$^3$LL) accuracy in the infinite top quark mass approximation. We discuss the theoretical uncertainties that are stemming from the ambiguities in resummation. The final results, ready for phenomenological applications, are reported as N$^3$LL matched to N$^3$LO (N$^3$LO+N$^3$LL), which are also augmented with the known full $m_t$-dependent NLO calculation.

The remainder of the paper is organised as follows. We describe our theoretical framework in section \ref{sec:theoryframe}. Our results are reported in section \ref{sec:results}. The conclusions are drawn in section \ref{sec:conclusions}. The concrete expressions of the universal coefficients appearing in the N$^3$LL resummation for $gg\to hh$ can be found in appendix \ref{app:A}.

\section{Theoretical framework\label{sec:theoryframe}}

\subsection{General structure}
For a general hadronic process with hadrons $h_1$ and $h_2$ collide to produce a heavy colourless final state with the invariant mass square $M^2$, the differential invariant mass distribution in QCD improved parton model takes the form of a convoluted integral:
\begin{equation}\label{eq:partonmodel}
    M^2 \dfrac{d}{d M^2}\sigma_{h_1h_2}(s,M^2) = \tau \sum_{a,b=q,\bar q,g} \int_\tau^1 \frac{dz}{z} \phi_{ab}\left(\frac{\tau}{z},\mu_F^2\right)~ \Delta_{ab}  \left(z,M^2,\mu_F^2\right)\,.
\end{equation}
 Here, $s$ is the square of the hadronic center-of-mass energy, $\mu_F$ is the factorisation scale and the dimensionless quantity $\tau$ is defined as $\tau\equiv \dfrac{M^2}{s}$. The partonic luminosity $\phi_{ab}$ is defined as the convolution of parton distribution functions (PDFs) $f_{a/h_1}$ and $f_{b/h_2}$:
\begin{equation}\label{eq:partonlum}
    \phi_{ab}(x,\mu_F^2) \equiv \int_x^1 \dfrac{dy}{y} f_{a/h_1}(y,\mu_F^2) f_{b/h_2}\left(\dfrac{x}{y},\mu_F^2\right).
\end{equation}
While the partonic coefficient function $\Delta_{ab} $ is perturbatively calculable in QCD, and is expanded in terms of $a_s \equiv \alpha_s/(4\pi)$ with $\alpha_s$ being the UV renormalised strong coupling constant.
In the collinear factorisation, the partonic coefficient function is expressed in terms of the bare partonic cross section $\hat{\sigma}_{ab}$ and the splitting kernels $\Gamma$ at the scale $\mu_F$:
\begin{eqnarray}
\label{eq:def-Cof-function}
\Delta_{ab}(z,M^2,\mu_F^2) 
&=&\sum_{c,d}{\Gamma^{-1}_{ca}(z,\mu_F^2,\mu^2,\epsilon)\otimes 
{\frac{1}{z}\hat{\sigma}_{cd}(z,M^2,\mu^2,\epsilon) }
\otimes \Gamma^{-1}_{db}(z,\mu_F^2,\mu^2,\epsilon)},
\end{eqnarray}
with all the initial collinear singularities encapsulated in $\Gamma$, in terms of the regularised Altarelli-Parisi (AP) splitting functions~\footnote{The lowest order is $\Gamma^{-1}_{ab}=\delta_{ab}$ where $\delta_{ab}$ is the Kronecker delta function.}.
The $\mu_F$ dependence, in principle, should cancel order by order at the hadronic level after convoluting $\Delta_{ab}(z,M^2,\mu_F^2) $ with the partonic luminosity.
$\mu$ in eq.~\eqref{eq:def-Cof-function} is the regularisation scale in the dimensional regularisation with $d=4-2\epsilon$ spacetime dimensions. The symbol $\otimes$ refers to the convolution operator defined for arbitrary functions $f_i(x),i=1,2,\ldots,n$ as:
\begin{eqnarray}
\label{convolution}
        f_1(z) \otimes f_2(z) \otimes \cdot \cdot \cdot \otimes f_n(z) 
\equiv \prod_{i=1}^n \Bigg(\int_0^1 dx_i f_i(x_i)\Bigg)~ \delta(z - x_1 x_2 \cdot\cdot \cdot x_n ) \,. 
\end{eqnarray}
 The bare partonic cross section $\hat{\sigma}_{ab}$ can be expanded in $\hat{a}_s\equiv\frac{\hat{\alpha}_s}{4\pi}$ with the bare strong coupling constant $\hat{\alpha}_s$:
\begin{equation}
       \hat{\sigma}_{ab}(z,M^2,\mu^2,\epsilon) = \hat{a}_s^b S_{\epsilon}^b  \left[\hat{\sigma}^{(0)}_{ab}\delta(1-z)+\sum_{k=1}{\hat{a}_s^k ~\hat{\sigma}_{ab}^{(k)}}\right]\,,\label{eq:asexpand4partonicxs}
    \end{equation}
where $b$ is the power of $\hat{a}_s$ at the lowest order and $S_\epsilon\equiv (4\pi e^{-\gamma_E})^\epsilon$ in the spacetime dimension $d=4-2\epsilon$ with $\gamma_E$ being the Euler-Mascheroni constant. Beyond the lowest order in $\hat{a}_s$, the coefficient $\hat{\sigma}_{ab}^{(k)}$ involves UV and IR divergences. The UV renormalisation procedure removes the UV singularities and introduces a dependence on the renormalisation scale $\mu_R$. In the $\overline{\rm MS}$ scheme, the renormalised coupling constant $a_s$ is related to the bare coupling $\hat{a}_s$ via
\begin{align}
\label{eq:bBH-ashatANDas}
&{\hat{a}_s} S_{\epsilon}  = \left( \frac{\mu^{2}}{\mu_R^{2}}
  \right)^{-\epsilon}  Z_{\alpha_{s}}(\mu_R^2) a_s(\mu_R^2)\,,
\end{align}
which satisfies the renormalisation group running $\frac{d}{d\ln{\mu_R^2}}a_s=\beta(a_s)-\epsilon a_s$
in $d=4-2\epsilon$ dimensions. 
The beta function ensures the following series expansion
\begin{align}
    \beta(a_s)=-a_s\sum_{k=0}{\beta_k a_s^{k+1}}
\end{align}
with
$ \beta_0=\frac{11}{3}C_A-\frac{4}{3}T_Fn_q, \beta_1=\frac{1}{3}\left(34C_A^2-12C_FT_Fn_q-20C_AT_Fn_q\right)$, etc. 
Here, $C_F=\frac{4}{3}$ and $C_A=3$ are Casimir constants, $T_F=\frac{1}{2}$ and $n_q$ is the number of massless quark flavours.
The renormalisation group invariance of the bare coupling constant $\frac{d}{d\ln{\mu_R^2}}\ln{\hat{a}_s}=0$ gives us $\beta(a_s(\mu_R^2))=-a_s(\mu_R^2)\frac{d}{d\ln{\mu_R^2}}\ln{Z_{\alpha_s}(\mu_R^2)}$ with
\begin{align}
\label{eq:bBH-Zas}
Z_{\alpha_{s}}(\mu_R^2) &= 1+ a_s(\mu_R^2) \left[-\frac{\beta_0}{\ep} \right]
           + a_s^2(\mu_R^2) \left[\frac{\beta_0^2}{\ep^2 }-\frac{\beta_1}{2\ep}  \right]
           + a_s^3(\mu_R^2) \left[-\frac{ \beta_0^3}{ \ep^3} +\frac{7}{6
            \ep^2} \beta_0 \beta_1 - \frac{\beta_2}{3 \ep}   \right]
\nonumber\\
          &+ a_s^4(\mu_R^2) \left[\frac{\beta_{0}^{4}}{\ep^{4}} -
            \frac{23}{12\ep^{3}}\beta_{0}^{2}\beta_{1} +
            \frac{1}{\ep^{2}}\left(\frac{3}{8}\beta_{1}^{2} +
            \frac{5}{6}\beta_{0}\beta_{2}\right) -
            \frac{1}{4\ep}\beta_{3} \right]+\mathcal{O}(a_s^5)\,.
\end{align}
If the amplitude depends on other bare couplings like the Higgs Yukawa couplings or the other effective couplings, it is understood that the similar UV renormalisation should be carried out. For simplicity, we ignore the dependence on other couplings in the following context.

The $\alpha_s$ series expansion given in eq.(\ref{eq:asexpand4partonicxs}) is usually referring as the fixed order perturbative calculations. However, the perturbative expansion has limitations in its applicability if the coefficients at each order involve large logarithms. Such cases could arise from the soft-gluon emissions. Despite the cancellation of soft singularities between virtual and real radiative contributions, soft gluon effects can yield potentially large logarithms in the kinematic configurations where high imbalance between real and virtual gluons happens. In such cases, the threshold resummation is suitable to be performed by resuming these large logarithms to all orders in $\alpha_s$, which is exactly the subject of the present paper.

\subsection{Soft-virtual limit}
    
Let us define the threshold limit in terms of the partonic variable $z \equiv \frac{M^2}{\hat{s}} \to 1$, with $\hat{s}$ being the square of the partonic center of mass energy. As the name implies, in this phase space region,  almost all the incoming partonic center of mass energy goes to produce the Born-like final
state, and the residual energy is only sufficient to generate soft gluons. 

In this limit, the partonic coefficient function can be organised into two parts:
\begin{align}
\label{eq:bBH-PartsOfDelta}
 &{\Delta}_{ab}(z, M^{2}, \mu_F^2) = {\Delta}^{
  \text{SV}}_{ab} (z, M^{2},  \mu_F^2) + {\Delta}^{
  \text{Reg}}_{ab} (z, M^{2},  \mu_F^2) \,.
\end{align}
In such a decomposition, the term $\Delta^{
   \text{SV}}_{ab} (z, M^{2},  \mu_F^2) $ involves only the singular terms when $z=1$ with the distributions of the following forms:
   \begin{equation}
      \left\{ \delta(1-z), {\cal D}_i(z) \equiv  \left[\frac{\ln^i{(1-z)}}{1-z}\right]_+ \right\}\,,
   \end{equation}
where $\delta(1-z)$ is the Dirac delta function and ${\cal D}_i(z)$ is a plus distribution. The latter is defined as that for any test function $f(z)$, there is
\begin{eqnarray}
\int_{0}^{1}{dz {\cal D}_i(z) f(z)}&=&\int_0^1{dz \frac{\ln^i{(1-z)}}{1-z}\left[f(z)-f(1)\right]}.
\end{eqnarray}
The superscript $\text{SV}$ denotes the soft-virtual part, accounting for the real emission corrections at the threshold limit and the pure virtual corrections. The remaining regular term $\Delta^{\text{Reg}}_{ab} (z, M^{2}, \mu_F^2)$ contains either $\ln^i{(1-z)}$ terms or the polynomials of $(1-z)$. Although the logarithms $\ln^i{(1-z)}$ could also be large, they are subdominant by a factor of $(1-z)$ compared to the soft-virtual contribution. In the following discussion, we ignore these subleading power terms.  

Since we focus on soft-virtual part in the context, only the flavour diagonal terms of $\Gamma$ in eq.\eqref{eq:def-Cof-function} contribute. Hence, the parton flavour sum in eq.\eqref{eq:def-Cof-function} can be safely omitted, giving rise to a simpler relation:
 \begin{eqnarray}
\label{eq:def-SV-Cof-function}
\Delta^{\text{SV}}_{ab}(z,M^2,\mu_F^2) 
&=&\Gamma_{aa}^{-1}(z,\mu_F^2,\mu^2,\epsilon)\otimes 
{1 \over z}\hat{\sigma}_{ab}^{\text{SV}}(z,M^2,\mu^2,\epsilon)
\otimes \Gamma_{bb}^{-1}(z,\mu_F^2,\mu^2,\epsilon).
\end{eqnarray}
Again, the superscript $\text{SV}$ indicates that we only keep those terms which give rise to $\delta(1-z)$ and ${\cal D}_i$.
The soft-virtual partonic cross section can be further factorised into the soft and hard (or virtual) parts. In this respect, the partonic coefficient function further takes the form:
\begin{align}\label{eq:chap1-SV-master-formula}
    \Delta^{\text{SV}}_{ab}(z,M^2,\mu_F^2)=&~\hat{a}_s^b S_{\epsilon}^b \hat{\sigma}^{(0)}_{ab}~|{\cal F}_{ab}(M^2,\mu^2,\epsilon)|^2~\delta(1-z)\otimes 
    \nonumber \\ &
    \left(\mathcal{S}_{ab}(z,M^2,\mu^2,\epsilon)
\otimes \Gamma_{aa}^{-1}(z,\mu_F^2,\mu^2,\epsilon)
\otimes \Gamma_{bb}^{-1}(z,\mu_F^2,\mu^2,\epsilon)\right)\,.
\end{align}
 The form factor ${\cal F}_{ab}$ captures pure loop corrections and starts from $1$ as the lowest term. The implicit dependence of the form factor on external particle momenta, colours and spins is left understood.
 The soft function $\mathcal{S}_{ab}$ embeds the real soft-gluon corrections. Each piece on the right-hand side of eq.~\eqref{eq:chap1-SV-master-formula} is not UV renormalised yet. Hence they involve UV divergences, which should be removed by the UV renormalisation. ${\cal F}_{ab}$ and $\mathcal{S}_{ab}$ contain IR poles as well. $\Delta^{\text{SV}}_{ab}$, on the other hand, is both UV and IR finite, the details of which will be discussed shortly.
 
 In dimensional regularisation, the IR divergences in the form factor stemming from the soft and collinear loop-momentum modes can be renormalised as~\footnote{In the following, if an object has the dependence on the renormalisation scale $\mu_R$, it means we have carried out the UV renormalisation for the coupling constant $\alpha_s$ unless we explicitly write out its dependence on $\hat{a}_s$ or $\hat{\alpha}_s$. After the UV renormalisation, there is no explicit $\mu$ dependence anymore, because the regularisation scale in the dimensional regularisation is always in the form of $(\mu^2)^{\epsilon}$ for each loop-momentum integration.}:
\begin{align}
\label{eq:M-Mfin}
      \hat a_s^bS_{\epsilon}^b\hat{\sigma}^{(0)} |{\cal F}_{ab}\left(M^2,\mu^2,\epsilon\right)|^2
     =\lim_{\epsilon \rightarrow 0} |Z^{\text{IR}}_{ab}(M^2,\mu_R^2,\epsilon)|^2 H_{ab}\left(M^2,\mu_R^2\right) 
      \,.
\end{align}
A few comments are in order. The coupling constant is renormalised on the right-hand side. Both $Z^{\text{IR}}_{ab}$ and $H_{ab}$ are perturbatively expanded in terms of the renormalised coupling constant $a_s$. $Z^{\text{IR}}_{ab}$ factorises all the IR divergences coming from the virtual amplitudes. The hard function $H_{ab}$ is finite and depends on the Born kinematics, such that the lowest order equals to the Born amplitude squared. From the IR factorisation of virtual amplitudes, it is well-known that 
$Z^{\text{IR}}_{ab}$ is a universal quantity~\cite{Catani:1998bh,Sterman:2002qn,Ravindran:2004mb,Moch:2005tm,Aybat:2006wq,Aybat:2006mz,Becher:2009cu,Gardi:2009qi,Almelid:2015jia,Almelid:2017qju}. For colourless particles production, it is insensitive to the hard process under study and only depends on if the process is initial quark or gluon induced. More specifically, the IR singularities of the virtual contribution are solely determined by the nature of the initial particles. This in turn leads to the fact that the combined contributions from the soft function $\mathcal{S}_{ab}$ and the kernels $\Gamma^{-1}$ must exhibit the same universal IR divergences. Absorbing  $Z^{\text{IR}}_{ab}$ into the second line of eq.\eqref{eq:chap1-SV-master-formula} and performing the coupling constant renormalisation at the scale $\mu_R$ yields:
\begin{align}\label{eq:partonicxsec}
    \Delta^{\text{SV}}_{ab}(z,M^2,\mu_F^2) =& H_{ab}(M^2,\mu_R^2) ~\delta(1-z)\otimes S_{\Gamma,ab}(z,M^2,\mu_F^2,\mu_R^2)\,
\end{align}
with
\begin{align}\label{SGamma}
    S_{\Gamma,ab}(z,M^2,\mu_F^2,\mu_R^2) \equiv& \lim_{\epsilon\to 0}|Z^{\text{IR}}_{ab}(M^2,\mu_R^2,\epsilon)|^2\delta(1-z) \otimes
    \nonumber \\ &
    \left[\mathcal{S}_{ab}(z,M^2,\mu_R^2,\epsilon)
    \otimes \Gamma_{aa}^{-1}(z,\mu_F^2,\mu_R^2,\epsilon) \otimes \Gamma_{bb}^{-1}(z,\mu_F^2,\mu_R^2,\epsilon)\right]\,.
\end{align}
Since the IR divergent part of  $\left(\mathcal{S}_{ab} \otimes \Gamma_{aa}^{-1} \otimes \Gamma_{bb}^{-1}\right)$ cancels against $|Z^{\text{IR}}_{ab}|^2$, both $H_{ab}$ and $S_{\Gamma,ab}$ are free of IR poles. The hard function $H_{ab}$ is obtained from the IR subtracted loop corrections~\footnote{Both $H_{ab}$ and $S_{\Gamma,ab}$ are IR subtraction scheme dependent due to the ambiguities in choosing $Z^{\text{IR}}_{ab}$. However, the combination of the two, $ \Delta^{\text{SV}}_{ab}$, is independent of the subtraction scheme.}. On the other hand, the structure of soft-collinear part $S_{\Gamma,ab}$ is obtained by computing real corrections and the AP splitting kernels in the soft limit. Alternatively, one can also predict the structure of soft-collinear part $S_{\Gamma,ab}$ by employing its universal behaviour. This leads to a framework to resum the large logarithms present at the threshold limit, which we discuss briefly in the following.

We begin with the ingredients in eq.\eqref{eq:chap1-SV-master-formula}. The structure of splitting kernel $\Gamma_{aa}$ is well-known. It satisfies the renormalisation group evolution equation with respect to $\mu_F$:
\begin{equation}\label{RGGam}
  \mu_F^2\frac{d}{d\mu_F^2}\Gamma_{aa}\big(z,\mu_F^2,\mu_{R}^2,\epsilon\big) = 
	\frac{1}{2} P_{aa} \big(z,a_s(\mu_F^2)\big)\otimes \Gamma_{aa}\big(z,\mu_F^2,\mu_{R}^2,\epsilon\big) + \mathcal{O}(1)\,, 
\end{equation}
where $P_{aa}$ is the regularised flavour diagonal (unpolarised) AP splitting function~\footnote{The three-loop unpolarised AP splitting functions are known in refs.~\cite{Moch:2004pa,Vogt:2004mw}. The non-singlet ones have been checked in ref.~\cite{Blumlein:2021enk} with a different approach.}. 
Note here that in the above equation, we have dropped the contributions from off-diagonal splitting functions since we focus on the soft-virtual contributions only. The diagonal splitting function at the production threshold takes the form:
\begin{align}
	P_{aa}\big(z,a_s(\mu_F^2)\big) = 2 B_{a}(a_s(\mu_F^2)) \delta(1-z) + 2 A_{a}(a_s(\mu_F^2)) {\cal D}_0(z) + \mathcal{O}(1) \,,
\end{align}
in terms of the light-like cusp $A_{a}$~\footnote{The 4-loop cusp anomalous dimension has been computed analytically in refs.~\cite{Henn:2019swt,vonManteuffel:2020vjv}.} and the virtual $B_{a}$~\footnote{The complete 4-loop virtual anomalous dimensions are firstly known in ref.~\cite{Das:2019btv} for quarks and in ref.~\cite{Das:2020adl} for gluons.} anomalous dimensions, which depend only on the nature of incoming particle $a$. In particular, we notice that $A_a(a_s)=\sum_{k=1}{a_s^k A_a^{(k)}}=a_s 4C_a+\mathcal{O}(a_s^2)$ and $B_{a}(a_s)=\sum_{k=1}{a_s^k B_{a}^{(k)}}=\left\{\begin{array}{ll}a_s~\beta_0+\mathcal{O}(a_s^2), & a=g\\
a_s 3C_F+\mathcal{O}(a_s^2), & a=q,\bar{q}\\
\end{array}\right.$, where $C_q=C_{\bar{q}}=C_F$ and $C_g=C_A$ are Casimir constants. 
The solution of
eq.~\eqref{RGGam} is found to be 
\begin{align}\label{eq:splittingkernalform}
   \Gamma_{aa}\big(z,\mu_F^2,\mu^2,\epsilon\big)  = \delta(1-z) + \sum_{k=1} \hat a_s^k S_\epsilon^k \left(\frac{\mu_F^2}{\mu^2}\right)^{-k\epsilon} \Gamma_{aa}^{(k)}\big(z,\epsilon\big)\,.
\end{align}
The expressions of $\Gamma_{aa}^{(k)}\big(z,\epsilon\big)$ up to $k=4$ in the $\rm \overline{MS}$ scheme can be found in eq.~(33) of ref.~\cite{Ravindran:2005vv} by replacing $\varepsilon$ with $-2\epsilon$. In particular, $\Gamma_{aa}^{(1)}\big(z,\epsilon\big)=-\frac{1}{\epsilon}\left(A_a^{(1)}{\cal D}_0(z)+B_a^{(1)}\delta(1-z)+\mathcal{O}(1)\right)$.

The structure of the soft function ${\cal S}_{ab}$ could be determined from the behaviour of the Sudakov form factor through its evolution under the momentum transfer $M^2$, which leads to its exponentiation~\cite{Mueller:1979ih,Collins:1980ih,Sen:1981sd}, and the renormalisation group evolution of the splitting kernels $\Gamma_{aa}$ and $\Gamma_{bb}$.
The universal factorisation of IR singularities guarantees that the soft function could be expressed in terms of a first order differential equation. From eq.\eqref{eq:chap1-SV-master-formula}, we can derive
\begin{align}\label{KGsoft}
    M^2\frac{d}{dM^2}\ln {\cal S}_{ab} &=
    M^2\frac{d}{dM^2} \ln \Delta_{ab}^{\text{SV}}-M^2\frac{d}{dM^2} \left(2\ln {\cal F}_{ab}+\ln{\hat{\sigma}_{ab}^{(0)}}\right) \delta(1-z)
    \nonumber \\ &
    =-2M^2\frac{d}{dM^2}\ln Z^{\text{IR}}_{ab} \delta(1-z)+ M^2\frac{d}{dM^2}\ln S_{\Gamma,ab}
    \nonumber \\ &
    =-2M^2\frac{d}{dM^2}\ln Z^{\text{IR}}_{ab} \delta(1-z)+ M^2\frac{d}{dM^2}\ln S_{ab}^{\rm fin}
    \nonumber \\ 
    & =  \overline{K}_{ab} (\hat a_s, \frac{\mu_R^2}{\mu^2},z,\epsilon) + \overline{G}_{ab}(\hat a_s,\frac{M^2}{\mu_R^2},\frac{\mu_R^2}{\mu^2},z,\epsilon)\,,
\end{align}
where we have defined the finite remainder of the soft function as $S_{ab}^{\rm fin}$. The coefficient
$\overline{K}_{ab}$ is resulting from the IR pole part, which only involves the soft divergences in the real contribution. It can be related to $Z^{\text{IR}}_{ab}$ as following:
\begin{align}
   \overline{K}_{ab} (\hat{a}_s, \frac{\mu_R^2}{\mu^2},z,\epsilon) =  -2M^2\frac{d}{dM^2}\ln Z^{\text{IR}}_{ab} \delta(1-z)\,.
\end{align}
Expanding them in the bare coupling $\hat{a}_s$ gives:
\begin{align}
    \overline{K}_{ab}(\hat a_s, \frac{\mu_R^2}{\mu^2},z,\epsilon) =  \sum\limits_{k=1} {\hat a}_s^k
              S_{\epsilon}^{k} \left(
              \frac{\mu_R^2}{\mu^2} \right)^{-k\epsilon}
              \overline{K}^{(k)}_{ab}(\epsilon) \delta(1-z)=\sum_{k=1}{Z_{\alpha_s}^k(\mu_R^2)a_s^k(\mu_R^2)\overline{K}^{(k)}_{ab}(\epsilon) \delta(1-z)}\,.
\end{align}
The coefficient $\overline{K}^{(k)}_{ab}(\epsilon)$ solely depends on the cusp ($A_{a}$) anomalous dimension and the beta function. The relationship is given in eq.(35) of ref.~\cite{Ravindran:2006cg} by replacing $\varepsilon$ with $-2\epsilon$ and $A^I_k$ with $A_a^{(k)}$.
Further, the renormalisation group invariance of ${\cal S}_{ab}$ (i.e., $\frac{d}{d\ln \mu_R^2}{\cal S}_{ab}(z,M^2,\mu^2,\epsilon)=0$) yields:
\begin{align}\label{eq:KGInv}
\frac{d}{d\ln \mu_R^2} \overline{G}_{ab} &= 
-\frac{d}{d\ln \mu_R^2}\overline{K}_{ab} =\epsilon\sum_{k=1}{k\hat{a}_s^k S_\epsilon^k\left(\frac{\mu_R^2}{\mu^2}\right)^{-k\epsilon}\overline{K}_{ab}^{(k)}(\epsilon)}\delta(1-z)
\nonumber\\
&=\epsilon\sum_{k=1}{kZ^k_{\alpha_s}(\mu_R^2)a_s^k(\mu_R^2)\overline{K}_{ab}^{(k)}(\epsilon)}\delta(1-z)
= -A_{a}\delta(1-z)\,.
\end{align}
This is obtained from the peculiar structure of $ \ln Z_{ab}^{\rm IR}(M^2,\mu_R^2,\epsilon) $ with $Z_{ab}^{\rm IR}(M^2,\mu_R^2,\epsilon)=\sum_{k=1}{a_s^k(\mu_R^2) Z_{ab}^{{\rm IR},(k)}(M^2,\mu_R^2,\epsilon)}$ at every order in the perturbative expansion:
\begin{eqnarray}
    \frac{d}{d\ln \mu_R^2} \ln Z^{{\rm IR},(k)}_{ab}(M^2,\mu_R^2,\epsilon)  &=& -\frac{1}{2}A_a^{(k)} \ln \left(\dfrac{-M^2}{\mu_R^2}\right) + B_a^{(k)} + \frac{f_a^{(k)}}{2}\nonumber\\
    &=&-\frac{1}{2}A_a^{(k)} \ln \left(\dfrac{-M^2}{\mu_R^2}\right)-\gamma_a^{{\rm collinear},(k)}\,.
\end{eqnarray}
$f_a(a_s)=\sum_{k=1}{a_s^k f_a^{(k)}}$ is called soft/eikonal anomalous dimension~\footnote{The four loop quark soft anomalous dimension is given in ref.~\cite{Das:2019btv}, while the 4-loop gluon counterpart is obtained in ref.~\cite{Das:2020adl} from the quark one with the conjecture that they satisfy the generalised Casimir scaling rule~\cite{Moch:2018wjh}.} with $f_a^{(1)}=0$. Up to three loops in QCD, the cusp and soft anomalous dimension satisfies a Casimir scaling relationship given by $\frac{A_g}{A_q}=\frac{f_g}{f_q}=\frac{C_g}{C_q}=\frac{C_A}{C_F}$. The collinear anomalous dimension, which is analytically known up to 4 loops~\cite{Chakraborty:2022yan}, can be related to the virtual and soft anomalous dimensions via $\gamma_{a}^{\rm collinear}=\sum_{k=1}{a_s^k\gamma_a^{{\rm collinear},(k)}}=-B_a-\frac{f_a}{2}$. From eq.\eqref{eq:KGInv}, $\overline{G}_{ab}$ takes the form:
\begin{align}
&\overline{G}_{ab} \Big( {\hat a}_s, \frac{M^2}{\mu_R^2},
  \frac{\mu_R^2}{\mu^2},z,\epsilon\Big) = \underbrace{\sum\limits_{k=1} \hat{a}_s^k S_\epsilon^k \left(\frac{M^2}{\mu^2}\right)^{-k\epsilon} \overline G^{(k)}_{ab}(z,\epsilon)}_{=\overline{G}_{ab} \Big( {\hat a}_s, 1,
  \frac{M^2}{\mu^2},z,\epsilon\Big)} -\underbrace{\delta(1-z)
         \int_{\frac{M^2}{\mu_R^2}}^1
         \frac{d\lambda^2}{\lambda^2}
  A_a\left(a_s(\lambda^2\mu_R^2)\right)}_{=\overline{K}_{ab}(\hat{a}_s,\frac{\mu_R^2}{\mu^2},z,\epsilon)-\overline{K}_{ab}(\hat{a}_s,\frac{M^2}{\mu^2},z,\epsilon)}\,,
\end{align}
where the first term on the right-hand side is the boundary term at $\mu_R=M$, whose explicit form is determined from the soft limit of real corrections.

In light of the differential equation \eqref{KGsoft} and the renormalisation group invariance eq.(\ref{eq:KGInv}), the solution takes the form~\footnote{The $M^2$ independent term in $\ln {\cal S}_{ab}(z, M^2,\mu^2, \epsilon)$ from the general solution of the first-order differential equation \eqref{KGsoft} is vanishing.}:
\begin{align}
    \ln {\cal S}_{ab}(z, M^2,\mu^2, \epsilon) &= { -\sum_{k=1}\hat{a}_s^kS_\epsilon^{k}\Big(\frac{M^2}{\mu^2}\Big)^{-k\epsilon}~\frac{1}{k\epsilon}
\left(\overline K^{(k)}_{ab}(\epsilon) \delta(1-z) + \overline{G}_{ab}^{(k)}(z,\epsilon)\right)}\,.
\end{align}
Alternatively, with an understanding on the structure of $\overline{G}_{ab}^{(k)}(z,\epsilon)$ from the explicit results
\begin{align}
  \overline G_{ab}^{(k)}(z,\epsilon) =  \big((1-z)^2\big)^{-k\epsilon} \Big[\frac{-2k\epsilon}{1-z}\overline G_{ab}^{(k)}(\epsilon) +  \left(\frac{-2k\epsilon}{1-z}\right)_+ \overline K_{ab}^{(k)}(\epsilon)\Big]\,,
\end{align}
one can also formulate a better functional form for $\ln {\cal S}_{ab}(z, M^2,\mu^2, \epsilon)$:
\begin{eqnarray}\label{PhiSV1}
\ln {\cal S}_{ab}(z, M^2,\mu^2,\epsilon) &=&2 \sum_{k=1}\hat{a}_s^kS_\epsilon^{k}\Big(\frac{M^2(1-z)^2}{\mu^2}\Big)^{-k\epsilon} 
 \frac{1}{1-z}\left( \overline K_{ab}^{(k)}(\epsilon) + \overline G_{ab}^{(k)}(\epsilon)\right)\,,
\end{eqnarray}
which satisfies all the aforementioned differential equations.
Here we have used the relationship
\begin{eqnarray}\label{chap1-dist}
    \frac{1}{1-z} \left[ (1-z)^2\right]^{-k\epsilon}&=&\frac{1}{-2k\epsilon}\delta(1-z) +  \left[\frac{1}{1-z} \left( (1-z)^2\right)^{-k\epsilon}\right]_+\nonumber\\
    &=&\frac{1}{-2k\epsilon}\delta(1-z) +  \sum_{j=0}{\frac{(-2k\epsilon)^j}{j!}{\cal D}_j(z)}\,.
\end{eqnarray}
The exact form of $\overline{G}_{ab}^{(k)}(\epsilon)$ is given in eq.(37) of ref.~\cite{Ravindran:2006cg} by replacing $\varepsilon$ with $-2\epsilon$ and identifying the subscript $P$ as $S$ (therefore $\delta_{P,SJ}=0$). In particular, we have $\overline{G}_{ab}^{(1)}(\epsilon)=\epsilon \frac{3}{2}\zeta_2 A_a^{(1)}+\mathcal{O}(\epsilon^2)$. Equation~\eqref{PhiSV1} obviously shows that when one takes $\mu\sim\mathcal{O}(M(1-z))$, the soft function is free of any large logarithms, which should take the initial scale of its renormalisation group evolution. The proper initial scale is just at the order around $M(1-z)$, thus suffering from the ambiguities of the initial scale choices. In eq.~\eqref{PhiSV1}, the terms $\Big(\dfrac{M^2(1-z)^2}{\mu^2 }\Big)^{-\epsilon}$ and $\dfrac{1}{(1-z)}$ are from two body phase space and the real matrix element squared respectively. 
For more details on this structure and for explicit results, we refer readers to  refs.~\cite{Ravindran:2005vv, Ravindran:2006cg, Ahmed:2020nci}.
 The form of the soft function as given in eq.~\eqref{PhiSV1} is same or universal for any arbitrary production process of colourless final states. This is evident since the soft enhancements only depend on the nature of the external coloured particles, i.e., the initial partons. It is also interesting to find that ${\cal S}_{ab}$ satisfies maximally non-abelian property up to third order in the strong coupling constant~\cite{Ravindran:2005vv,Ravindran:2006cg,Li:2014bfa,Catani:2014uta}, by which the quark and gluon lines are related by the Casimir constants in the adjoint and fundamental representations. In other words, we have $\dfrac{{\cal S}_{gg}}{{\cal S}_{q\bar q}} = \dfrac{C_A}{C_F}$. However, whether the validity of this property holds beyond third order with the so-called generalised Casimir scaling relation~\cite{Moch:2018wjh} needs to be addressed in future. In the present paper, we only need the knowledge of the 3-loop soft function ${\cal S}_{gg}$.

\subsection{Threshold resummation}

Threshold logarithmic corrections dominate when the partonic scaling variable $z$ approaches to unity. This is manifested in eq.~\eqref{PhiSV1} in terms of $
    \left\{\delta(1-z), {\cal D}_i(z)\right\}$ which is evident by using the relation eq.~\eqref{chap1-dist}.
In this kinematic limit, the large logarithm $\ln(1-z)$ multiplying with the small coupling $a_s$ produces ${\cal O}(1)$ terms, i.e., $ a_s \ln (1-z) \sim \mathcal{O}(1) $\,. This potentially spoils the perturbative convergence if we truncate at a fixed order in $a_s$. In such a case, the threshold resummation procedure~\cite{Sterman:1986aj,Catani:1989ne} provides an alternative perturbative expansion. 

In order to construct the resummation framework, we employ the structure of soft function ${\cal S}_{ab}$, AP splitting kernel and the master equation \eqref{eq:partonicxsec} that we have discussed in the last subsection. Using the relation eq.~\eqref{chap1-dist}, the soft function can be decomposed into the Dirac delta part and the plus distribution part:
\begin{align}
    \ln {\cal S}_{ab} = \ln {\cal S}_{ab, \delta} ~\delta(1-z) + \ln {\cal S}_{ab,{\cal D}}\,,
\end{align}
where $\ln {\cal S}_{ab, \delta}$ is proportional to $\delta(1-z)$ and $\ln {\cal S}_{{ab},{\cal D}}$ involves only plus distributions of the form ${\cal D}_i$. 
Rearranging eq.~\eqref{PhiSV1}, we get:
\begin{eqnarray}
\ln {\cal S}_{ab,\delta}=
- \sum_{k=1} \hat{a}_s^kS_{\epsilon}^k
\left({M^2 \over \mu^2}\right)^{-k\epsilon}~
\frac{1}{k\epsilon}\left(\overline{K}_{ab}^{(k)}(\ep) + \overline{G}_{ab}^{(k)} (\ep) \right)\,.
\label{Sdelta}
\end{eqnarray}
The above expression involves $\epsilon$ poles and finite terms, where the former will cancel against the IR poles of the virtual correction $Z_{ab}^{\rm IR}$ and of the splitting kernels $\Gamma^{-1}_{aa},\Gamma^{-1}_{bb}$. Similarly, the distribution part of the soft function is given by:
\begin{align}
\ln {\cal S}_{ab,{\cal D}}= &
~2\sum_{k=1} \hat{a}_s^k S_{\epsilon}^k
\left({M^2 \over \mu^2}\right)^{-k\epsilon} \left(\frac{(1-z)^{-2k\epsilon}}{1-z}\right)_+
~\left(\overline K_{ab}^{(k)}(\ep) + \overline G_{ab}^{(k)} (\ep) \right)
\nonumber \\ 
=&   \left(\frac{2}{1-z} \sum_{k=1} a_s^k\left(M^2(1-z)^2\right) Z_{\alpha_s}^k\left(M^2(1-z)^2\right)
~\overline G_{ab}^{(k)} (\ep)\right)_+
\nonumber \\ &
+ \left(\frac{2}{1-z}\sum_{k=1}\hat{a}_s^k S_{\epsilon}^k 
\left(\frac{\mu_F^2}{\mu^2}\right)^{-k\epsilon}\left[\left({M^2 (1-z)^2 \over \mu_F^2}\right)^{-k\epsilon} - 1\right] 
\overline K_{ab}^{(k)}(\ep)\right)_+
\nonumber \\ &
+\left({2 \over 1-z}\right)_+ ~\sum_{k=1}
\hat{a}_s^k S_{\ep}^k\left({\mu_F^2 \over \mu^2}\right)^{-k\ep}~
\overline K^{(k)}_{ab}(\ep)\,.
\end{align}
This gives us an integral representation for $\ln {\cal S}_{ab,{\cal D}}$:
\begin{align}
\ln {\cal S}_{ab,{\cal D}}=&~2\Bigg( {1 \over 1-z} \Bigg\{
\int_{\mu_F^2}^{M^2 (1-z)^{2}} {d \lambda^2 \over \lambda^2}
A_a \left(a_s(\lambda^2)\right) + \overline {\cal G}_{ab} \left(
a_s\left(M^2 (1-z)^{2}\right),\ep\right)\Bigg\} \Bigg)_+
\nonumber\\&
+\left({2 \over 1-z}\right)_+ ~\sum_{k=1}
\hat{a}_s^k S_{\ep}^k\left({\mu_F^2 \over \mu^2}\right)^{-k \ep}~
\overline K^{(k)}_{ab}(\ep)\,,
\label{SD}
\end{align}
where $\overline{{\cal G}}_{ab} \left(a_s(\mu_R^2),\ep\right)$ is
\begin{align}
    \overline{{\cal G}}_{ab} \left(
a_s\left(\mu_R^2\right),\ep\right) \equiv \sum_{k=1}a_s^k(\mu_R^2) Z^k_{\alpha_s}(\mu_R^2)\overline{G}_{ab}^{(k)}(\ep)\,.
\end{align}
In eq.~\eqref{SD}, we have used the relation
\begin{align}
\label{eq:App-SolveG-2}
 \sum\limits_{k=1}{\hat a}_s^k ~S_{\epsilon}^k ~\left(
  \frac{\mu_F^2}{\mu^2}\right)^{-k\epsilon} \left[ \left(
  \frac{M^2}{\mu_F^2} \right)^{-k\epsilon} -1 \right] {\overline K}_{ab}^{(k)}(\epsilon) = \int^{M^2}_{\mu_F^2} \frac{d\lambda^2}{\lambda^2} A_a\left(a_s(\lambda^2)\right) 
 \,.
\end{align}
The IR poles in the second line in eq.~\eqref{SD} cancel against the poles of the ${\cal D}_0$ term of the splitting kernels exactly. The finite remnants of the splitting kernels depend on $\ln (\mu_F^2/\mu_R^2)$ when expressing eq.~\eqref{eq:splittingkernalform} in the renormalised coupling constant $a_s$ at the scale $\mu_R$.

Convoluting the soft function with the AP splitting kernels and $Z_{ab}^{\rm IR}$ produces the IR finite function $S_{\Gamma,ab}$ defined in eq.~\eqref{SGamma}:
\begin{align}
\ln S_{\Gamma,ab}(z,M^2,\mu_F^2,& \mu_R^2)=
2 \ln Z_{ab}^{\rm IR} \delta(1-z)+2 \ln {\cal S}_{ab,\delta}\delta(1-z) + 2  \ln {\cal S}_{ab,{\cal D}} -\ln \Gamma_{aa}- \ln \Gamma_{bb}\,.
\end{align}
As mentioned above, the poles of $\ln {\cal S}_{ab,\delta}$ and $\ln {\cal S}_{ab, {\cal D}}$ cancel against the respective terms from $Z_{ab}^{\rm IR}$ , $\Gamma_{aa}$ and $\Gamma_{bb}$. Decomposing the splitting kernels into the Dirac delta and plus distribution parts:
\begin{align}
    \ln \Gamma_{aa} = \ln \Gamma_{aa, \delta} ~\delta(1-z) + \ln \Gamma_{aa,{\cal D}}\,,
\end{align}
and further expressing the soft function and the splitting kernels into the singular and finite terms:
\begin{align}
    \ln {\cal S}_{ab,\{\delta,{\cal D}\}} = \ln {\cal S}_{ab,\{\delta,{\cal D}\}}^{\rm sing} + \ln {\cal S}_{ab,\{\delta,{\cal D}\}}^{\rm fin}\,,
    \nonumber\\
    \ln \Gamma_{aa,\{\delta,{\cal D}\}} = \ln \Gamma_{aa,\{\delta,{\cal D}\}}^{\rm sing} + \ln \Gamma_{cc,\{\delta,{\cal D}\}}^{\rm fin}\,,
\end{align}
$S_{\Gamma,ab}$ becomes~\footnote{Note that since $S_{\Gamma,ab}$ is IR finite, we can take $\epsilon=0$ directly for all the quantities appearing in it.}:
\begin{align}
\ln S_{\Gamma,ab}&(z,M^2,\mu_F^2,\mu_R^2)=
\left( 
 2\ln {\cal S}_{ab,\delta}^{\rm fin} - \ln \Gamma_{aa,\delta}^{\rm fin}-\ln \Gamma_{bb,\delta}^{\rm fin}\right) \delta(1-z) + (2\ln {\cal S}_{ab,{\cal D}}^{\rm fin} -  \ln \Gamma_{aa,{\cal D}}^{\rm fin} - \ln \Gamma_{bb,{\cal D}}^{\rm fin})
\nonumber \\
= & \left( 
 2\ln {\cal S}_{ab,\delta}^{\rm fin} - \ln \Gamma_{aa,\delta}^{\rm fin}-\ln \Gamma_{bb,\delta}^{\rm fin}\right) \delta(1-z)
 \nonumber\\ &
+~\Bigg( {2 \over 1-z} \Bigg\{
\int_{\mu_F^2}^{M^2 (1-z)^{2}} {d \lambda^2 \over \lambda^2}
A_a \left(a_s(\lambda^2)\right) + \overline {\cal G}_{ab} \left(
a_s\left(M^2 (1-z)^{2}\right)\right)\Bigg\} \Bigg)_+ \,,
\end{align}
where we have defined
\begin{align}
    \overline {\cal G}_{ab} \left(
a_s\left(M^2 (1-z)^{2}\right)\right)\equiv \lim_{\ep\to 0}{\overline {\cal G}_{ab} \left(
a_s\left(M^2 (1-z)^{2}\right),\ep\right)}\,.
\end{align}
Here, the first line comprises of the finite part of $ \ln {\cal S}_{ab,\delta}$ and the finite remnants arising from the $\delta(1-z)$ terms of the splitting kernels. 

Finally, taking into account the finite hard function $H_{ab}$, as in eq.~\eqref{eq:partonicxsec}, it gives us the partonic coefficient function:
\begin{align}\label{Delta}
      \Delta^{\rm{SV}}_{ab}(z,M^2,& \mu_F^2) = ~g_{0,ab} \Big(M^2,\mu_F^2,\mu_R^2\Big) ~\delta(1-z) ~\otimes
     \\ &
     {\cal P} \exp\Bigg( {2 \over 1-z} \Bigg\{
\int_{\mu_F^2}^{M^2 (1-z)^{2}} {d \lambda^2 \over \lambda^2}
A_a \left(a_s(\lambda^2)\right) + \overline {\cal G}_{ab} \left(
a_s\left(M^2 (1-z)^{2}\right)\right)\Bigg\} \Bigg)_+ \,,
\nonumber
\end{align}
where
\begin{align}
    g_{0,ab}\Big(M^2,\mu_F^2,\mu_R^2\Big) \equiv  H_{ab}\Big(M^2,\mu_R^2\Big) \times \exp
    \left( 
 2\ln {\cal S}_{ab,\delta}^{\rm fin} - \ln \Gamma_{aa,\delta}^{\rm fin} -\ln \Gamma_{bb,\delta}^{\rm fin}\right)\,.
\end{align}
 The symbol $\mathcal{P}$ refers to the operator for the ``path-ordered exponential", which has the following expansion for an arbitrary function $f(z)$:
\begin{eqnarray}\label{chap2-Cexp}
\mathcal{P}e^{f(z)} = \delta(1-z) + \frac{1}{1!}f(z) + \frac{1}{2!}(f(z)\otimes f(z)) + \cdots \,.
\end{eqnarray}
 The second line in eq.~\eqref{Delta} involves enhanced logarithms in the limit $z\rightarrow 1$ which have been resummed. Since this term is in the form of convolutions, it is convenient to solve it in the Mellin $N$-space, where all the convolutions turn into normal products. The Mellin transform of a function $f(z)$ is generally defined as:
\begin{equation}
    {\cal M}[f(z)](N) \equiv \int_0^1 dz ~z^{N-1} f(z)\,.
\end{equation}
The threshold limit $z\rightarrow 1$ corresponds to $N\rightarrow \infty$ in the Mellin $N$-space. Note that ${\cal M}[\delta(1-z)] = 1$ and ${\cal M}[{\cal D}_k(z)] = \frac{(-1)^{k+1}}{(k+1)}\ln^{k+1} N + {\cal O}(\ln^{k} N)$. In the Mellin space, eq.~\eqref{Delta} reads as:
\begin{align}\label{resummation}
    \Delta^{\rm res}_{ab}(N,M^2,\mu_F^2)&  \equiv \int_0^1 dz ~z^{N-1} \Delta^{\rm{SV}}_{ab}(z,M^2,\mu_F^2)
    \nonumber \\ &
=~g_{0,ab}\Big(M^2,\mu_F^2,\mu_R^2\Big)\times 
 \\ &
~~~~~\exp \left[
    2\int_0^1 dz \frac{z^{N-1}-1}{1-z}\left(
\int_{\mu_F^2}^{M^2 (1-z)^{2}} {d \lambda^2 \over \lambda^2}
A_a \left(a_s(\lambda^2)\right) + \overline {\cal G}_{ab} \left(
a_s\left(M^2 (1-z)^{2}\right)\right) \right) \right]\,.
    \nonumber \\
\nonumber
\end{align}
Solving the above integral in the Mellin space with an expansion over the resummation parameter $\omega \equiv 2 \beta_0 a_s(\mu_R^2) \ln N $ results into:
\begin{align}
    \Delta_{ab}^{\rm res}(N,M^2,\mu_F^2) &= g_{0,ab}(M^2,\mu_F^2,\mu_R^2)\times \nonumber\\
    &~\exp \left(\tilde{C}_{0,ab} (a_s(\mu_R^2)) + g_{1,ab} (\omega)\ln N + \sum_{k=2}{a_s^{k-2}(\mu_R^2)~g_{k,ab} (\omega)}\right)\,.
\end{align}
As been noticed from the above equation, the exponent after the Mellin transform has an additional $N$-independent coefficient, $\tilde{C}_{0,ab}$, arising from the ${\cal O}(1)$ terms of ${\cal M}[{\cal D}_k(z)]$. This coefficient is in general depending on the Riemann zeta function $\zeta_n$ and the Euler-Mascheroni constant $\gamma_E$. Furthermore, the $N$-dependent functions $g_{k,ab}, k\ge 1$ are all-order coefficients expressed in terms of $\omega\sim \mathcal{O}(1)$. These coefficients satisfy the condition that they vanish in the limit $N\rightarrow 1$. Their dependencies on $\ln \frac{M^2}{\mu_R^2}$ and $\ln \frac{M^2}{\mu_F^2}$ are also implicitly understood.
Taking out all the $N$-independent terms from the exponent gives us:
\begin{align}\label{DeltaN}
    \Delta_{ab}^{\rm res}(N,M^2,\mu_F^2)  = \tilde{g}_{0,ab}(M^2,\mu_F^2,\mu_R^2)~\exp \Big( g_{1,ab} (\omega)\ln N + \sum_{k=2}{a_s^{k-2}(\mu_R^2)g_{k,ab}(\omega)}\Big)\,.
\end{align}
The total $N$-independent prefactor is
\begin{align}\label{g0t}
\tilde{g}_{0,ab} (M^2,\mu_F^2,\mu_R^2)\equiv& ~g_{0,ab}(M^2,\mu_F^2,\mu_R^2) \exp\Big( \tilde C_{0,ab}(a_s(\mu_R^2))\Big)
 \nonumber \\
 = &~ H_{ab}\Big(M^2,\mu_R^2\Big) \times \exp\left( 
 2\ln {\cal S}_{ab,\delta}^{\rm fin} - \ln \Gamma_{aa,\delta}^{\rm fin} - \ln \Gamma_{bb,\delta}^{\rm fin} +\tilde C_{0,ab} \right)\,.
\end{align}
 As stated earlier, the resummation coefficients $g_{k,ab}(\omega)$ are universal and depend only on the nature of incoming partons. However, $\tilde{g}_{0,ab}$ is process dependent. It can be expanded in $a_s(\mu_R^2)$ as
\begin{equation}\label{g0tab}
  \tilde{g}_{0,ab}(M^2,\mu_F^2,\mu_R^2) = a_s^b(\mu_R^2)\sum_{k=0} a_s^k(\mu_R^2) ~\tilde{g}_{0,ab}^{(k)}(M^2,\mu_F^2,\mu_R^2)\,.
\end{equation}
Each term in the exponent in eq.~\eqref{DeltaN} has been resummed to all orders in $\alpha_s$. At the N$^{k}$LL accuracy, we should take into account both the logarithmic terms up to $g_{k+1,ab}(\omega)$ in the exponent and the $N$-independent terms up to $\tilde{g}_{0,ab}^{(k)}$ in $\tilde{g}_{0,ab}$. It is also implicitly understood that the hard function, as well as $g_{0,ab}$ and $\tilde{g}_{0,ab}$, may depend on the other underlying Born kinematic variables, which we have not yet explicitly written out. 

When the resummation calculations match to the fixed-order computations, we need to subtract the double counting between the two. At the $\NkLONkLL$ accuracy, the double counting is just the N$^k$LO soft-virtual term encoded in the partonic coefficient
function. In other words, it is the $a_s(\mu_R^2)$ series of $\Delta^{\rm res}_{ab}$~\footnote{Note that $\omega$ also depends on $a_s(\mu_R^2)$, and it should be expanded into the $a_s$ series too.} up to the $\mathcal{O}(a_s^{b+k})$ term.

\subsection{Ambiguities in resummation and resummation schemes\label{sec:resscheme}}

Up to a given logarithmic accuracy, the right-hand side of the resummation formalism eq.~(\ref{DeltaN}) is not uniquely defined. One has the freedom to choose which $N$-independent piece should be absorbed into the exponent. In other words, we can rewrite the resummed N$^k$LO partonic coefficient function in the Mellin space as
\begin{align}\label{resummationgeneral}
    \left.\Delta^{\rm res}_{ab}(N,M^2,\mu_F^2)\right|_{{\rm N}^k{\rm LL}}&  = \left[\tilde{g}_{0,ab}(M^2,\mu_F^2,\mu_R^2)\exp{\left(-\ddot{C}_{0,ab}(M^2,\mu_F^2,\mu_R^2)\right)}\right]_{{\rm N}^k{\rm LO}}\\ &
\times \exp{\left(\ddot{C}_{0,ab}(M^2,\mu_F^2,\mu_R^2)+g_{1,ab}(\omega)\ln N+\sum_{j=2}^{k+1}a_s^{j-2}(\mu_R^2)g_{j,ab}(\omega)\right)}\,,
\nonumber
\end{align}
where the function $\ddot{C}_{0,ab}(M^2,\mu_F^2,\mu_R^2)$ is an arbitrary $N$-independent function that can be expressed in an $a_s(\mu_R^2)$ series. The subscript N$^k$LO means that the whole expression in the bracket should be kept up to the first $k+1$ terms in the $a_s$ series. This essentially leads to the ambiguities in the threshold resummation formalism at a given logarithmic accuracy.~\footnote{Such resummation ambiguities have already been studied in the Drell-Yan process at N$^3$LL accuracy in ref.~\cite{Ajjath:2020rci}.}
In this subsection, we present a few concrete forms of $\ddot{C}_{0,ab}$ to discuss such ambiguities. First of all, in eq.~\eqref{DeltaN}, we resum the large logarithms of the form $\ln N$, with the resummation parameter given by $\omega$. However, it is observed that in the large $N$-expansion, $\ln N$ is always associated with the Euler-Mascheroni constant $\gamma_E$. Alternatively, one usually takes the resummation parameter as $\overline{\omega}\equiv 2\beta_0 a_s(\mu_R^2)\ln \overline{N}$ with $\overline{N}\equiv N e^{\gamma_E}$. In this paper, we consider four different resummation schemes as follows.
\begin{itemize}
\item $\ResNOne$ scheme : In this scheme, we have the N$^k$LL resummed partonic coefficient directly from eq.~\eqref{DeltaN}
\begin{align}\label{resummationN1}
    \left.\Delta^{{\rm res},\ResNOne}_{ab}(N,M^2,\mu_F^2)\right|_{{\rm N}^k{\rm LL}}&  = \left[\tilde{g}_{0,ab}(M^2,\mu_F^2,\mu_R^2)\right]_{{\rm N}^k{\rm LO}}\\ &
\times \exp{\left(g_{1,ab}(\omega)\ln N+\sum_{j=2}^{k+1}a_s^{j-2}(\mu_R^2)g_{j,ab}(\omega)\right)}\,.
\nonumber
\end{align}
\item $\ResNTwo$ scheme : In this scheme, we take $\ddot{C}_{0,ab}(M^2,\mu_F^2,\mu_R^2)=\tilde{C}_{0,ab,\zeta}\equiv\left.\lim_{\gamma_E\to 0}\tilde{C}_{0,ab}\right.$. In other words,  $\tilde{C}_{0,ab,\zeta}$ solely depends on the Riemann zeta function, and is defined as $\tilde{C}_{0,ab}$ by removing the $\gamma_E$-dependent terms. Similarly, we can define $\tilde{C}_{0,ab,\gamma_E}\equiv \tilde{C}_{0,ab}-\tilde{C}_{0,ab,\zeta}\propto \gamma_E$.
Then, the N$^k$LL resummed partonic coefficient becomes
\begin{align}\label{resummationN2}
    \left.\Delta^{{\rm res},\ResNTwo}_{ab}(N,M^2,\mu_F^2)\right|_{{\rm N}^k{\rm LL}} = \bigg[H_{ab}(M^2,\mu_R^2) &~\exp\Big(2\ln {\cal S}_{ab,\delta}^{\rm fin} - \ln \Gamma_{aa,\delta}^{\rm fin} - \ln \Gamma_{bb,\delta}^{\rm fin}  + \tilde C_{0,ab,\gamma_E}\Big) \bigg]_{{\rm N}^k{\rm LO}} 
    \nonumber\\
    \times \exp{\left[\left(\tilde C_{0,ab,\zeta} \right)_{{\rm N}^{k-1}{\rm LO}}\right]} &
     \exp{\left(g_{1,ab}(\omega)\ln N+\sum_{j=2}^{k+1}a_s^{j-2}(\mu_R^2)g_{j,ab}(\omega)\right)}\,.
\end{align}
In eq.~\eqref{resummationN2}, we use the subscript N$^{k-1}$LO in the exponent because the lowest order of $\tilde{C}_{0,ab,\zeta}$ is $\mathcal{O}(a_s)$. 
    \item $\ResNbOne$ scheme : In this scheme, we use the resummation parameter $\overline{\omega}$ instead of $\omega$. The coefficient is
    \begin{align}\label{resummationNb1}
    \left.\Delta^{{\rm res},\ResNbOne}_{ab}(N,M^2,\mu_F^2)\right|_{{\rm N}^k{\rm LL}}&  = \left[\overline{\tilde{g}}_{0,ab}(M^2,\mu_F^2,\mu_R^2)\right]_{{\rm N}^k{\rm LO}}\\ &
\times \exp{\left(\overline{g}_{1,ab}(\overline{\omega})\ln \overline{N}+\sum_{j=2}^{k+1}a_s^{j-2}(\mu_R^2)\overline{g}_{j,ab}(\overline{\omega})\right)}\,,
\nonumber
\end{align}
    where $\overline{g}_{j,ab}(\overline{\omega})=\left[\lim_{\gamma_E\to 0}g_{j,ab}(\omega)\right]_{\omega\to\overline{\omega}}$ when $j\geq 1$. This means that we have taken
    \begin{align}
    \ddot{C}_{0,ab}(M^2,\mu_F^2,\mu_R^2)&=\left(\overline{g}_{1,ab}(\overline{\omega})\ln \overline{N}+\sum_{j=2}^{k+1}a_s^{j-2}(\mu_R^2)\overline{g}_{j,ab}(\overline{\omega})\right)\nonumber\\
    &-\left(g_{1,ab}(\omega)\ln N+\sum_{j=2}^{k+1}a_s^{j-2}(\mu_R^2)g_{j,ab}(\omega)\right)\,.
    \end{align}
    It can be proven that $\ddot{C}_{0,ab}(M^2,\mu_F^2,\mu_R^2)$ is indeed independent of $N$. Thus, $\overline{\tilde{g}}_{0,ab}(M^2,\mu_F^2,\mu_R^2)=\tilde{g}_{0,ab}(M^2,\mu_F^2,\mu_R^2)\exp{\left(-\ddot{C}_{0,ab}(M^2,\mu_F^2,\mu_R^2)\right)}$.
    \item $\ResNbTwo$ scheme : In this scheme, we have the coefficient
\begin{align}\label{resummationNb2}
    \left.\Delta^{{\rm res},\ResNbTwo}_{ab}(N,M^2,\mu_F^2)\right|_{{\rm N}^k{\rm LL}} = \bigg[H_{ab}(M^2,\mu_R^2) &~\exp\Big(2\ln {\cal S}_{ab,\delta}^{\rm fin} - \ln \Gamma_{aa,\delta}^{\rm fin} - \ln \Gamma_{bb,\delta}^{\rm fin}  \Big) \bigg]_{{\rm N}^k{\rm LO}} 
    \nonumber\\
    \times \exp{\left[\left(\tilde C_{0,ab,\zeta} \right)_{{\rm N}^{k-1}{\rm LO}}\right]} &
\exp{\left(\overline{g}_{1,ab}(\overline{\omega})\ln \overline{N}+\sum_{j=2}^{k+1}a_s^{j-2}(\mu_R^2)\overline{g}_{j,ab}(\overline{\omega})\right)}\,.
\end{align}
Notice that in this scheme the $\gamma_E$-dependent piece $\tilde{C}_{0,ab,\gamma_E}$ has been absorbed into the transforms of $N\to\overline{N}$ and $\omega\to\overline{\omega}$ in the exponent.
\end{itemize}

\subsection{N$^3$LL resummation for Higgs boson pair production}

With the general formalism for the threshold resummation of any colourless final state, we are now in the position to say something specifically for the Higgs pair production in ggF. We begin with the master formula for the threshold resummation given in eq.~\eqref{DeltaN}. The quantities in the exponent given by $g_{k,gg}(\omega)$ are known up to $k\leq 4$,~\footnote{The explicit result of $\overline{g}_{4,gg}(\overline{\omega})$ in eq.~\eqref{resummationNb1} can also be found in appendix G of ref.~\cite{Ahmed:2020nci} by replacing $\omega$ with $\overline{\omega}$. Their general expressions for both quarks and gluons are in principle also given, e.g., in ref.~\cite{Ajjath:2019neu}.} which were first applied for the single Higgs production up to the N$^3$LL accuracy~\cite{Moch:2005ky}. These expressions can be fully recycled for the Higgs boson pair process. Their expressions up to the fourth order are provided in appendix \ref{app:A} for the completeness.

The $N$-independent coefficient $\tilde{g}_{0,gg}$ is not universal, whose process dependence is only stemming from the hard function $H_{gg}$. The universal $N$-independent contributions from the soft function and the splitting kernels could be found in appendices D and E of ref.~\cite{Ahmed:2020nci}. For the convenience of reader, we also provide their explicit results for the ggF processes up to the three loops in appendix \ref{app:A}. Combining them together with the hard function returns $\tilde{g}_{0,gg}^{(k)}$ appearing in eq.~\eqref{g0tab}.

The hard function $H_{gg}$ up to 3 loops for the N$^3$LL threshold resummation of the di-Higgs process in the infinite top quark mass approximation has been documented in appendix A of ref.~\cite{Chen:2019fhs}. For the sake of completeness, we still briefly mentioned it here. The effective Lagrangian given in eq.(2.1) of ref.~\cite{Chen:2019fhs} contains two Wilson coefficients $C_h$ and $C_{hh}$. These coefficients up to ${\cal O}(a_s^4)$ are:
\begin{align}
     C_h =& -\frac{4 a_s}{3}    \bigg\{ 1 + 11 a_s +
     a_s^2   \bigg[ \frac{2777}{18}  - \frac{67}{6} n_q +
    L_t   \Big( 19 + \frac{16}{3} n_q \Big)  \bigg]  
        \nonumber \\& 
    + a_s^3   \bigg[ -  \frac{2761331}{648} +   \zeta_3 \frac{897943}{144}   +
        \Big( \frac{58723}{324} -   \zeta_3 \frac{ 110779}{216} \Big)  n_q -   \frac{6865 }{486}n_q^2
         \nonumber \\& 
      + L_t   \Big( \frac{4834}{9} +      \frac{ 2912}{27}n_q +    \frac{ 77 }{27}n_q^2 \Big)   
      +  L_t^2 \Big( 209 + 46  n_q -   \frac{32}{9}n_q^2 \Big)      \bigg]  \bigg\}\,,
  \\
  C_{hh} = &-\frac{4a_s}{3}    \bigg\{ 1 + 11  a_s +
     a_s^2   \bigg[ \frac{3197}{18} -\frac{1}{2} n_q + L_t 
     \Big( 19  + \frac{16}{3 } n_q \Big)  \bigg]
         \nonumber \\&
    + a_s^3   \bigg[ -  \frac{2633363}{648} +    \zeta_3  \frac{897943}{144}   +    \Big( \frac{110611}{324} -     \zeta_3  \frac{110779}{216} \Big) n_q-      \frac{4093}{486}n_q^2
    \nonumber \\&
         + L_t \Big( \frac{11902}{9} +   \frac{7496}{27}  n_q -   \frac{307}{27} n_q^2\Big)    +
        L_t^2 \Big( 209 + 46  n_q -  \frac{32}{9}  n_q^2 \Big)  
         \bigg]  \bigg\}
     \end{align}
with $L_t \equiv \ln(\mu_R^2/m_t^2)$ and $m_t$ is the on-shell top quark mass. It is important to notice that the Wilson coefficients are at least ${\cal O}(a_s)$. Thus, we can organise the amplitude square into three parts dubbed as class-$a,b,c$, which feature $2,3,$ and $4$ insertions of the effective vertices respectively. This means that the hard function can be written as
\begin{equation}
    H_{gg} =  H_{gg}^a +  H_{gg}^b +  H_{gg}^c,
\end{equation}
whose expressions in terms of the amplitudes are given in eq.~(A.10) of ref.~\cite{Chen:2019fhs}. The class-$a$ hard function $H_{gg}^a$ has the similar topologies as the single Higgs process, known up to four loops~\cite{Lee:2022nhh}. The class-$b$ and -$c$ hard functions need the knowledge of other topologies. In eq.~(A.6) of ref.~\cite{Chen:2019fhs}, besides the trivial typo ${\cal M}_i^{B,(1)}$ that should be ${\cal M}_i^{B,(1),fin}$, ${\cal M}_i^{B,(2),fin}$ should be the finite $2$-loop amplitude defined in eq.~(2.24) of ref.~\cite{Banerjee:2018lfq} times $\frac{9C_h^2}{16a_s^2}$. The analytic expression of ${\cal M}_i^{B,(2),fin}$ is too lengthy to be directly used in the numerical code. In our calculation, for the practical usage, we have to generate the numerical grids in {\sc\small Mathematica} with 100-digit precision and with the help of the package {\sc\small PolyLogTools}~\cite{Duhr:2019tlz} first. By knowing the hard function, the $N$-independent coefficient $\tilde{g}_{0,gg}$ in eq.~(\ref{g0t}) becomes:
\begin{align}\label{g0g}
\tilde{g}_{0,gg}(m_{hh}^2,\mu_F^2,\mu_R^2)=&~(H_{gg}^a + H_{gg}^b + H_{gg}^c) \exp\Big(2\ln {\cal S}^{\rm fin}_{gg,\delta}-2\ln \Gamma^{\rm fin}_{gg,\delta} + \tilde{C}_{0,gg}\Big)
\nonumber \\ 
 = &~ a_s^2(\mu_R^2)~\sum_{k=0}{a_s^k(\mu_R^2)\tilde{g}^{(k)}_{0,gg}(m_{hh}^2,\mu_F^2,\mu_R^2)}\,,
\end{align}
where $m_{hh}$ is the invariant mass of the Higgs pair. The lowest orders of $H_{gg}^{a},H_{gg}^{b},H_{gg}^{c}$ contribute to $\tilde{g}^{(0)}_{0,gg},\tilde{g}^{(1)}_{0,gg},$ and $\tilde{g}^{(2)}_{0,gg}$ respectively.

We have carefully checked our implementation. For example, the hard function has been checked against ref.~\cite{Chen:2019fhs}, and its scale dependence satisfies the renormalisation group equation. In order to verify our numerical setup, we have also tested the N$^3$LL resummation calculation of the single Higgs process with ref.~\cite{Bonvini:2016frm}. In the $\ResNOne$ scheme, we find good agreements with ref.~\cite{deFlorian:2015moa} at the NNLL accuracy for the di-Higgs process.

\section{Results\label{sec:results}}

Before we present our results, we fix our setup for the numerical calculation. In order to recycle the N$^3$LO results that we have calculated before, we take the same setup as in ref.~\cite{Chen:2019fhs}. Namely, we take the Higgs mass $m_h=125$ GeV and the vacuum expectation value of the Higgs field $v=(\sqrt{2}G_F)^{-\frac{1}{2}}=246.2$ GeV. The top-quark pole mass $m_t=173$ GeV enters only into the Wilson coefficients $C_h$ and $C_{hh}$ of the effective Lagrangian. The same PDF {\tt PDF4LHC15\_nnlo\_30}~\cite{Butterworth:2015oua,Dulat:2015mca,Harland-Lang:2014zoa,NNPDF:2014otw}, along with its $\alpha_s$ renormalisation group running, provided by {\sc\small LHAPDF6}~\cite{Buckley:2014ana} will be used regardless of the $\alpha_s$ or logarithmic order we are considering. The default central scale is chosen as the half of the
invariant mass of the Higgs boson pair, i.e., $\mu_0=m_{hh}/2$. The scale uncertainty, for estimating the missing higher order terms, is evaluated through taking the envelope of the $7$-point variations of the factorisation scale $\mu_F$ and the renormalisation scale $\mu_R$ in the form of $\mu_{R/F}=\xi_{R/F}\mu_0$ with $\xi_{R/F}\in\ \{0.5,1,2\}$ after excluding the two extreme points $(\xi_R,\xi_F)=(0.5,2),(2,0.5)$.

\subsection{Impact of QCD corrections in the infinite top quark mass approximation}

We study the cross sections in the infinite top quark mass limit in this subsection. Both the inclusive cross sections and the Higgs pair invariant mass distributions will be reported.

\begin{figure*}[h]
\centering
\includegraphics[width=1\textwidth]{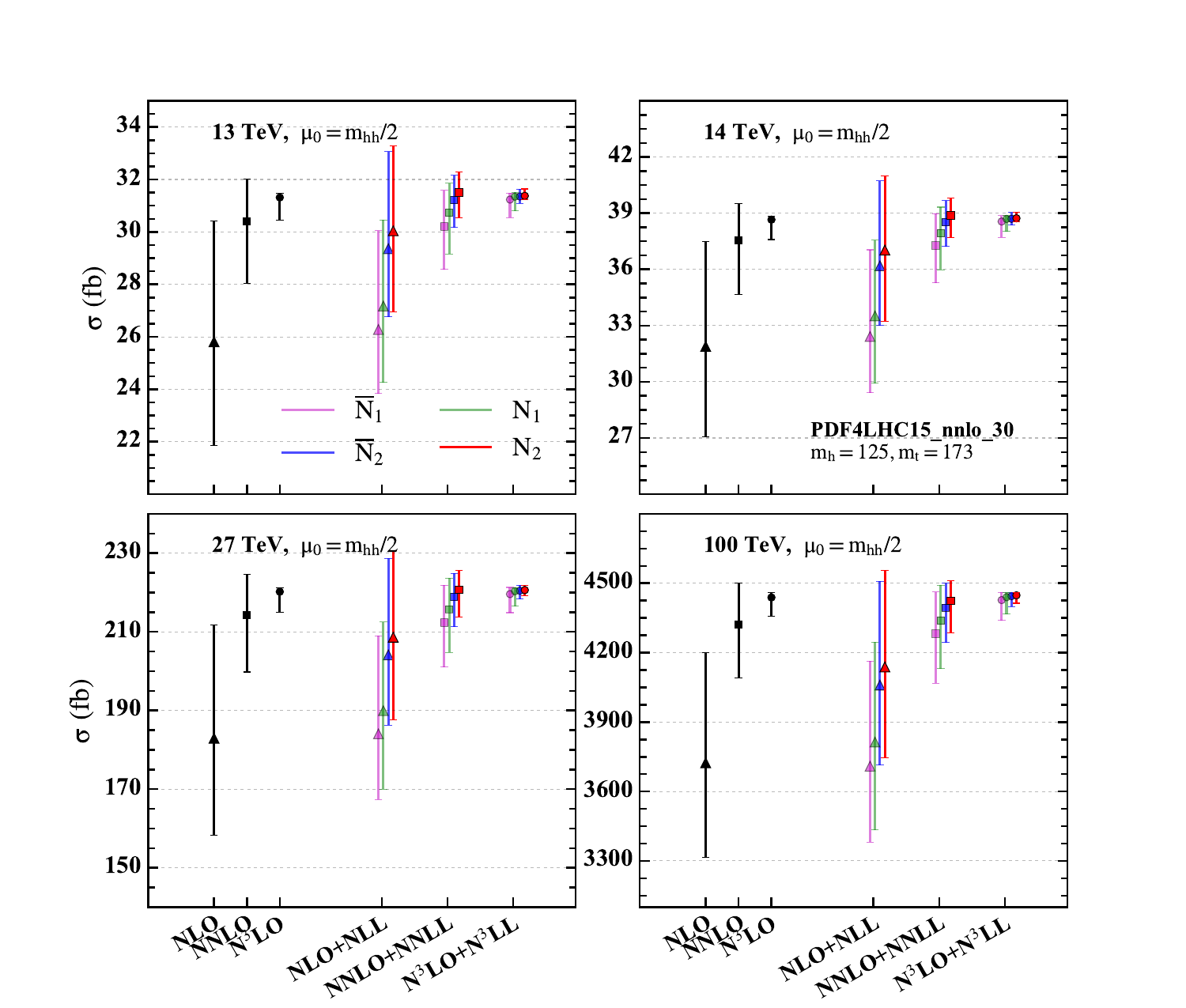}
\caption{The comparisons of the inclusive cross sections between the fixed order N$^k$LO calculations (black) and the threshold resummed N$^k$LO+N$^k$LL calculations (coloured) at four different energies $\sqrt{s}=13,14,27,100$ TeV. In each resummation calculation, we have shown four different results stemming from the resummation schemes. They are $\ResNOne$ (green), $\ResNTwo$ (red), $\ResNbOne$ (purple), and $\ResNbTwo$ (blue) schemes. The error bars are from 7-point scale variations.}
\label{fig:Verical_prescription}
\end{figure*}

Let us first focus on the inclusive cross sections that have been integrated with the whole phase space. As aforementioned, in the resummation calculations, we have the additional theoretical uncertainties stemming from the ambiguities in the resummation formalism discussed in subsection~\ref{sec:resscheme}. Figure~\ref{fig:Verical_prescription} reports the resummed results from NLO+NLL to N$^3$LO+N$^3$LL in the four typical center-of-mass energies covering the energy ranges of the LHC ($\sqrt{s}=13,14$ TeV) and future colliders ($\sqrt{s}=27,100$ TeV). Concrete numbers can also be read from table~\ref{tab:inclusivexs}. The four resummation schemes, dubbed as $\ResNOne,\ResNTwo,\ResNbOne,\ResNbTwo$, have been defined in subsection~\ref{sec:resscheme}. We have checked our NLL and NNLL calculations against those in ref.~\cite{deFlorian:2015moa}, where the latter only provides the $\ResNOne$ results. As anticipated, such a scheme dependence reduces from NLO+NLL to N$^3$LO+N$^3$LL. However, the different schemes give us both the different central values and error bars due to $7$-point scale variations. In particular, the differences are more significant between $\ResNbOne$ and $\ResNbTwo$ schemes than $\ResNbOne$ vs $\ResNOne$ or $\ResNbTwo$ vs $\ResNTwo$. The scale uncertainties are reduced only marginally from N$^k$LO to N$^k$LO+N$^k$LL in the $\ResNOne$ and $\ResNbOne$ schemes, while a factor two reductions have been observed in the $\ResNTwo$ and $\ResNbTwo$ schemes. For example, the N$^3$LO error bars are almost same after including the N$^3$LL resummation in the $\ResNbOne$ scheme. Besides the scale uncertainties, it is interesting to notice that the central values of N$^k$LO+N$^k$LL in the $\ResNTwo$ and $\ResNbTwo$ schemes are much closer to N$^{k+1}$LO than the other two. This potentially tells us that our best results N$^3$LO+N$^3$LL in the former two schemes can be extrapolated to predict the next order calculation, i.e., the unknown next-to-next-to-next-to-next-to-leading order (N$^4$LO). Although it seems that the $\ResNTwo$ and $\ResNbTwo$ schemes are equivalently good from the above perspectives, we decide to choose the $\ResNbTwo$ scheme in the following context as our best predictions, whose scale errors are more symmetric than the $\ResNTwo$ scheme. We want to stress that it is just our choice without any preference from first principles. An alternative attitude that is more conservative is to quote the theoretical uncertainties due to the scheme dependence instead of picking up one scheme. For instance, if we take the envelope of the N$^3$LO+N$^3$LL central values in the four schemes, we have $\sigma_{\rm N^3LO+N^3LL}=38.70\left(^{+0.85\%}_{-0.87\%}\right)_{\rm scale}\left(^{+0.08\%}_{-0.39\%}\right)_{\rm scheme}$ fb at $14$ TeV, where the second error due to the resummation scheme ambiguities is in anyway subdominant. 
Alternatively, if one takes the envelope of all the predictions in different schemes including the corresponding
scale uncertainties, the cross section at $14$ TeV   becomes  $\sigma_{\rm N^3LO+N^3LL}=38.70\left(^{+0.89\%}_{-2.6\%}\right)_{\rm scale+scheme}$ fb.
Nevertheless, according to the aforementioned arguments, we will stick to the $\ResNbTwo$ scheme for the following discussions.

\begin{table}[h] 
\begin{center}
\newcolumntype{P}[1]{>{\centering\arraybackslash}p{#1}}
{\renewcommand{\arraystretch}{1.5}
\begin{tabular}{|P{1.5cm}|P{1cm}|P{2cm}|P{2cm}|P{2cm}|P{2cm}|P{2cm}|}
\hline
     \multirow{2}{*}{ $\sqrt{s}$ [TeV]}
    & Order   
    &\multirow{2}{*}{N$^k$LO}  
    &\multicolumn{4}{c|}{N$^k$LO+N$^k$LL}\\ \cline{4-7} 
    &  $k$ & & $\ResNOne$ scheme &$\ResNTwo$ scheme &$\ResNbOne$ scheme & $\ResNbTwo$ scheme
    \\
 \hline
\multirow{4}{*}{13} & $0$ & $13.80^{+31\%}_{-22\%}$ & $16.01_{-23\%}^{+32\%}$ & $16.01_{-23\%}^{+32\%}$ & $21.02_{-24\%}^{+36\%}$ & $21.02_{-24\%}^{+36\%}$ \\
 & $1$ & $25.81^{+18\%}_{-15\%}$ & $  27.17_{-10.7\%}^{+12.1\%}$ & $30.04_{-10.3\%}^{+10.8\%}$  & $26.30_{-9.3\%}^{+14.4\%}$&  $29.36_{-8.8\%}^{+12.6\%}$\\
&$2$ &$30.41^{+5.3\%}_{-7.8\%}$  & $30.74_{-5.1\%}^{+3.7\%}$ &$31.51_{-3.0\%}^{+2.5\%}$ &$30.20_{-5.4\%}^{+4.6\%}$ & $31.21_{-3.3\%}^{+3.0\%}$\\
 &$3$ & $31.31^{+0.50\%}_{-2.8\%}$ & $31.34_{-1.7\%}^{+0.51\%}$ & $31.37_{-0.49\%}^{+0.84\%}$ &$31.23_{-2.2\%}^{+0.81\%}$ & $31.35_{-0.85\%}^{+0.88\%}$\\
  \hline
  \multirow{4}{*}{14}& $0$& $17.06^{+31\%}_{-22\%}$& $19.72_{-23\%}^{ +32\%}$ & $19.72_{-23\%}^{+32\%}$ &$25.80_{-24\%}^{+35\%}$&$25.80_{-24\%}^{+35\%}$\\
 & $1$ &$31.89^{+18\%}_{-15\%}$ & $33.52_{-10.7\%}^{+12\%}$ & $37.03_{-10.3\%}^{+10.7\%}$ &$32.42_{-9.3\%}^{+14.3\%}$& $36.19_{-8.8\%}^{+12.5\%}$ \\
 & $2$& $37.55^{+5.2\%}_{-7.6\%}$ & $37.93_{-5.1\%}^{+3.7\%}$ &$ 38.88_{-3.0\%}^{+2.4\%}$& $37.28_{-5.4\%}^{+4.5\%}$& $38.52_{-3.4\%}^{+3.0\%}$ \\
& $3$& $38.65^{+0.50\%}_{-2.7\%}$ & $38.69_{-1.7\%}^{+0.50\%}$ &$38.73_{-0.51\%}^{+0.81\%}$&$38.55_{ -2.2\%}^{+0.80\%}$& $38.70_{-0.87\%}^{+0.85\%}$ \\
 \hline
 \multirow{4}{*}{27} & $0$ & $98.22^{+26\%}_{-19\%}$&$ 110.6_{-20\%}^{+27\%}$ & $110.6_{-20\%}^{+27\%}$ & $141.1_{-21\%}^{+29\%}$ & $141.1_{-21\%}^{+29\%}$\\
 & $1$ &$183.0^{+16\%}_{-14\%}$&  $190.0_{-10.5\%}^{+11.9\%}$ &$208.6_{-10\%}^{+10.6\%}$  & $184.1_{-9.1\%}^{+13.5\%}$&  $204.2_{-8.8\%}^{+12.0\%}$\\
 &$2$ & $214.2^{+4.8\%}_{-6.7\%}$& $215.7_{-5.1\%} ^{+3.7\%}$ & $220.7_{-3.1\%}^{+2.2\%}$&$212.3_{-5.3\%}^{+4.5\%}$ & $218.9_{-3.4\%}^{+2.7\%}$\\
 &$3$ & $220.2^{+0.46\%}_{-2.3\%}$ & $220.3_{-1.7\%}^{+0.44\%}$ & $220.6_{-0.62\%}^{+0.57\%}$ &$219.6_{ -2.2\%}^{+0.77\%}$&$220.4_{-0.94\%}^{+0.63\%}$\\
  \hline
 \multirow{4}{*}{100} & $0$ & $2015^{+19\%}_{-15\%}$ &$2195_{-15\%}^{+19\%}$ & $2195_{-15\%}^{+19\%}$ & $2697_{-17\%}^{+21\%}$  & $2697_{-17\%}^{+21\%}$\\
 & $1$ &  $3724^{+13\%}_{-11\%}$ &$3813_{-10\%}^{+11.3\%} $  &$4138_{-9.5\%}^{+10.1\%}$  & $3709_{-8.8\%}^{+12.3\%}$& $4060_{-8.5\%}^{+11.1\%}$\\
 &$2$ &$4322^{+4.2\%}_{-5.3\%}$& $4338_{-4.8\%}^{+3.5\%}$ & $4424_{-3.1\%}^{+1.9\%}$ & $4283_{-5.0\%}^{+4.2\%}$  & $4394_{-3.4\%}^{+2.4\%}$\\
 &$3$ & $4439^{+0.51\%}_{-1.8\%}$ & $4440_{-1.6\%}^{+0.47\%} $& $4448_{-0.77\%}^{+0.26\%}$ &$4427_{ -2.0\%}^{+0.76\%}$& $4444_{-1.02\%}^{+0.35\%}$\\
 \hline
\end{tabular}}
\caption{Inclusive cross sections (in unit of fb) with the $m_t\to \infty$ approximation at both fixed order N$^k$LO and resummation N$^k$LO+N$^k$LL ($k=0,1,2,3$) in the center-of-mass energies $\sqrt{s}=13,14,27,100$ TeV. In the resummation, the results of four different resummation schemes have been shown. The quoted uncertainties are from the $7$-point scale variations with the central scale $\mu_0=\frac{m_{hh}}{2}$.}
\label{tab:inclusivexs}
\end{center}
\end{table}

\begin{figure*}[h]
\centering
\includegraphics[width=1\textwidth]{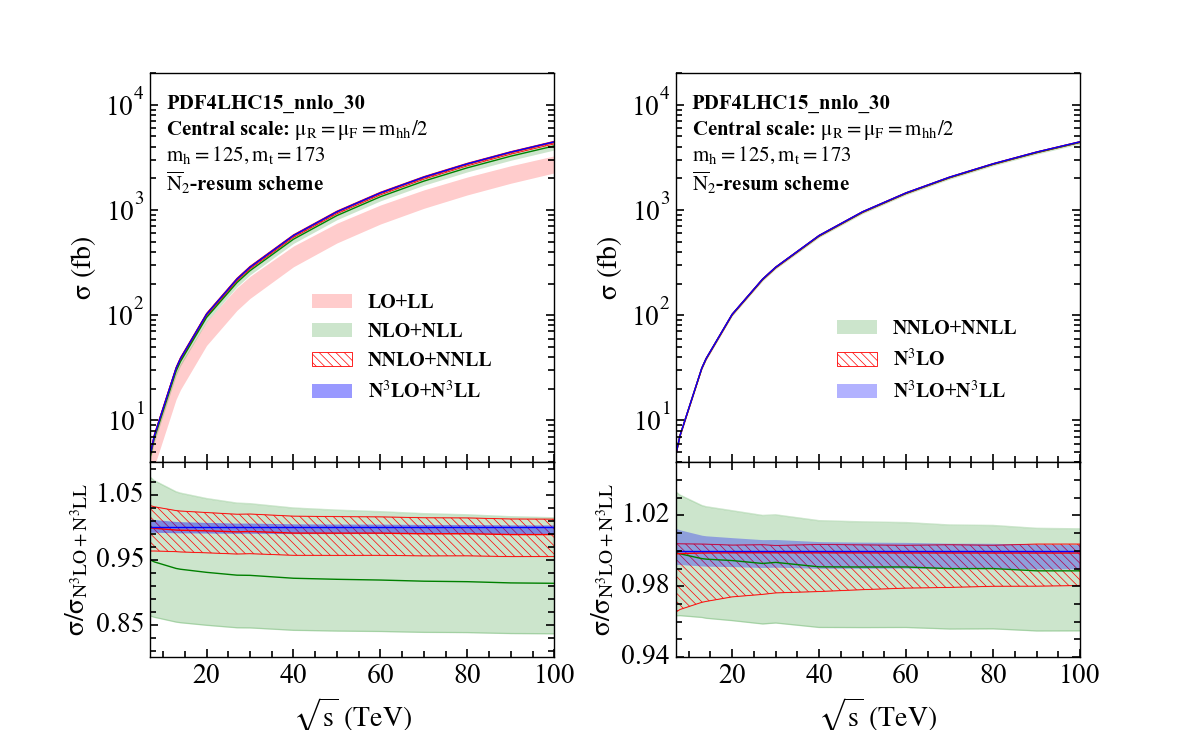}
\caption{Inclusive cross sections with respect to the center of mass energies $\sqrt{s}$ from LO+LL to N$^3$LO+N$^3$LL (left) and comparison study of the best result N$^3$LO+N$^3$LL against the previous best-known orders N$^3$LO and NNLO+NNLL (right). The error bands represent the 7-point scale uncertainties. In the lower panels, the ratios over the central value of N$^3$LO+N$^3$LL have been taken.}
\label{fig:Bandplot_best_NExp}
\end{figure*}

Table \ref{tab:inclusivexs} tells us N$^3$LO+N$^3$LL cross sections are enhanced only by the permille amounts from N$^3$LO. This corroborates and further consolidates our previous conclusion that the asymptotic convergence in $\alpha_s$ has reached at N$^3$LO~\cite{Chen:2019fhs,Chen:2019lzz}. The scale uncertainties of N$^3$LO+N$^3$LL are a factor $2$ smaller than N$^3$LO and almost a factor $4$ smaller than NNLO+NNLL, reaching to the sub-percent level now. The cross sections in a wider energy interval from $\sqrt{s}=7$ TeV to $\sqrt{s}=100$ TeV are displayed in figure~\ref{fig:Bandplot_best_NExp}. In the left plot, we compare various resummation orders from LO+LL to N$^3$LO+N$^3$LL. The error band of N$^k$LO+N$^k$LL, representing the $7$-point scale uncertainties, resides completely in the band of the previous order when $k\geq 2$.~\footnote{Without surprising, LO+LL clearly underestimates the missing higher order by using the manner of the scale variations.} Such an observation is analogous to the fixed order results~\cite{Chen:2019fhs,Chen:2019lzz}, with the exception that in the fixed order the N$^3$LO bands are outside the NLO counterparts. NLO+NLL increases LO+LL by $40\%$ ($50\%$) at $13$ ($100$) TeV. NNLO+NNLL further enhances NLO+NLL by $6\%$ and $8\%$ respectively at these two energies. Furthermore, N$^3$LO+N$^3$LL improves the NNLO+NNLL cross sections by only $0.4\%$ and $1\%$ at $\sqrt{s}=13$ and $100$ TeV respectively. The right plot of figure \ref{fig:Bandplot_best_NExp} compares our best prediction N$^3$LO+N$^3$LL to the previous state-of-the-art results known in the literature (NNLO+NNLL and N$^3$LO). Because the (partial) higher orders beyond N$^3$LO have been included in the NNLO+NNLL resummation calculation, strictly speaking, N$^3$LO does not supersede NNLO+NNLL. In contrast, N$^3$LO+N$^3$LL is superior to both of them. This is why the comparison in the right plot would be also interesting. From the ratios in the lower panel, the difference in the central values of N$^3$LO and N$^3$LO+N$^3$LL is almost invisible, while the difference between NNLO+NNLL and N$^3$LO+N$^3$LL is visible but within a percent. However, the N$^3$LO error band is completely enclosed in the NNLO+NNLL band, and the N$^3$LO+N$^3$LL error band largely lies in the N$^3$LO band. This again points to the good asymptotic perturbative convergence.

\begin{figure*}[htbp]
\centering
\includegraphics[width=1\textwidth]{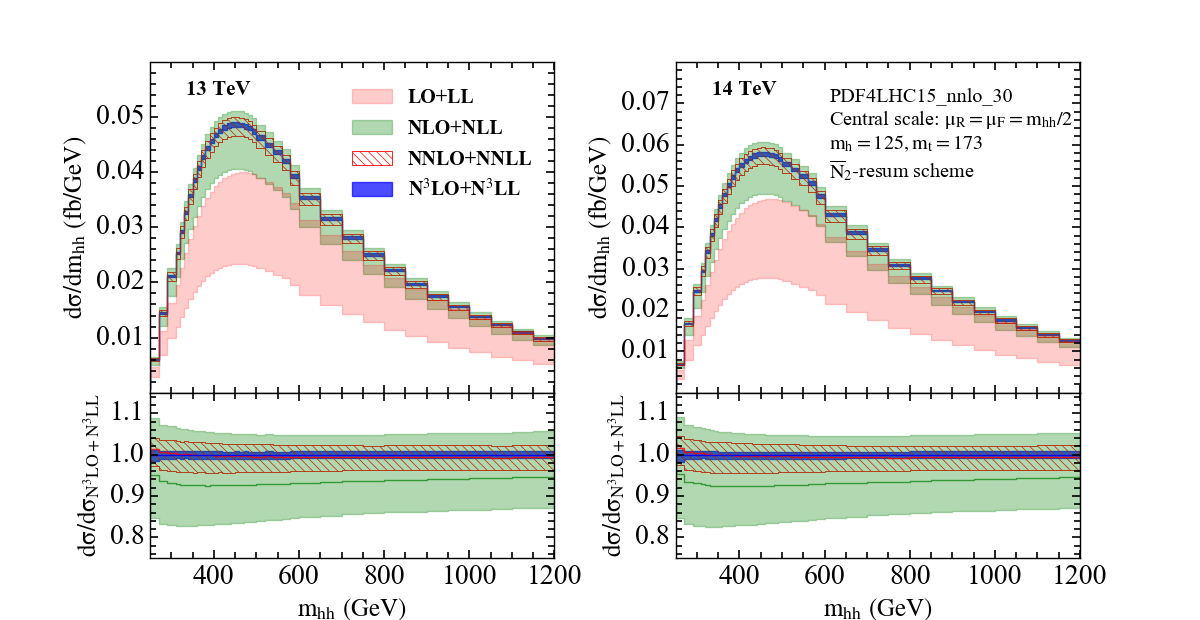}\\
\includegraphics[width=1\textwidth]{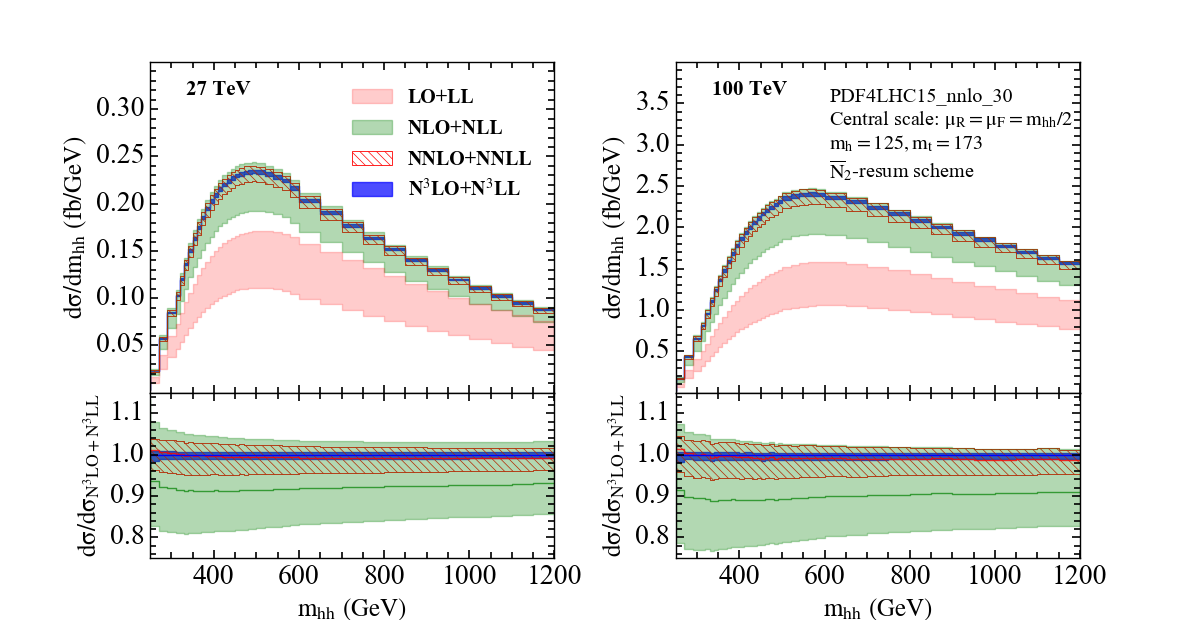}
\caption{The invariant mass differential cross sections $\frac{d\sigma}{dm_{hh}}$ from LO+LL to N$^3$LO+N$^3$LL at four different center-of-mass energies $\sqrt{s}=13,14,27,100$ TeV. The error bands represent the 7-point scale variations with the central scale $\mu_0=\frac{m_{hh}}{2}$. The pink, green, red hatched and blue bands correspond to the LL+LO, NLO+NLL, NNLO+NNLL and N$^3$LO+N$^3$LL predictions, respectively. The bottom panel shows the ratios to the N$^3$LO+N$^3$LL distribution.}
\label{fig:xs_vs_mhh_large_mt1}
\end{figure*}

\begin{figure*}[htbp]
\centering
\includegraphics[width=1\textwidth]{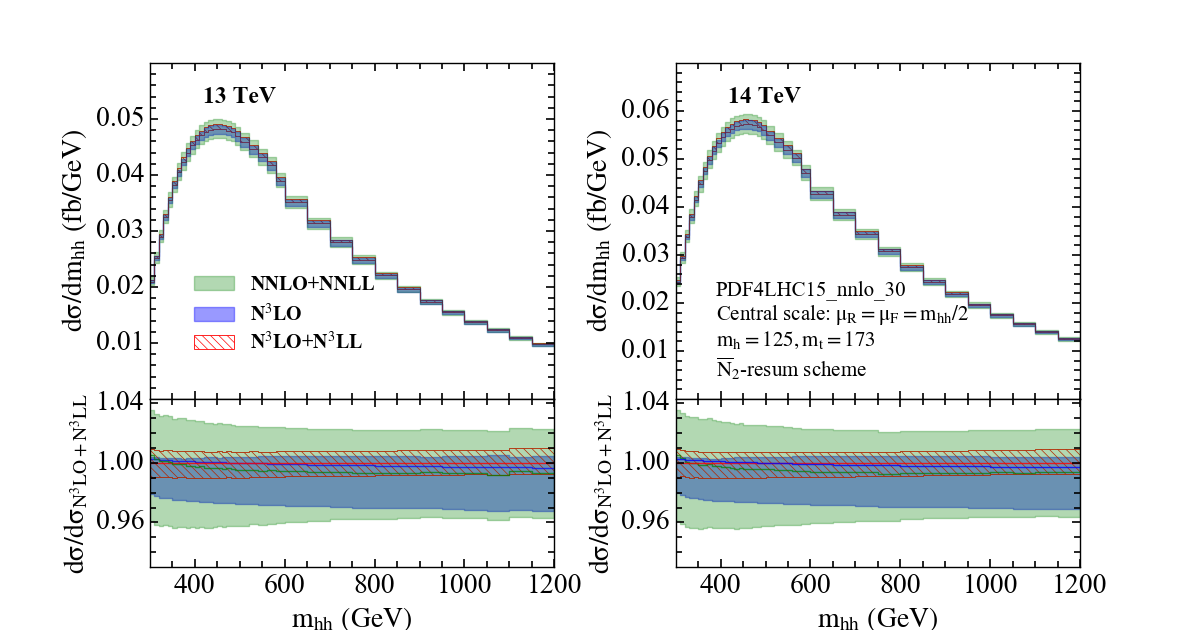}\\
\includegraphics[width=1\textwidth]{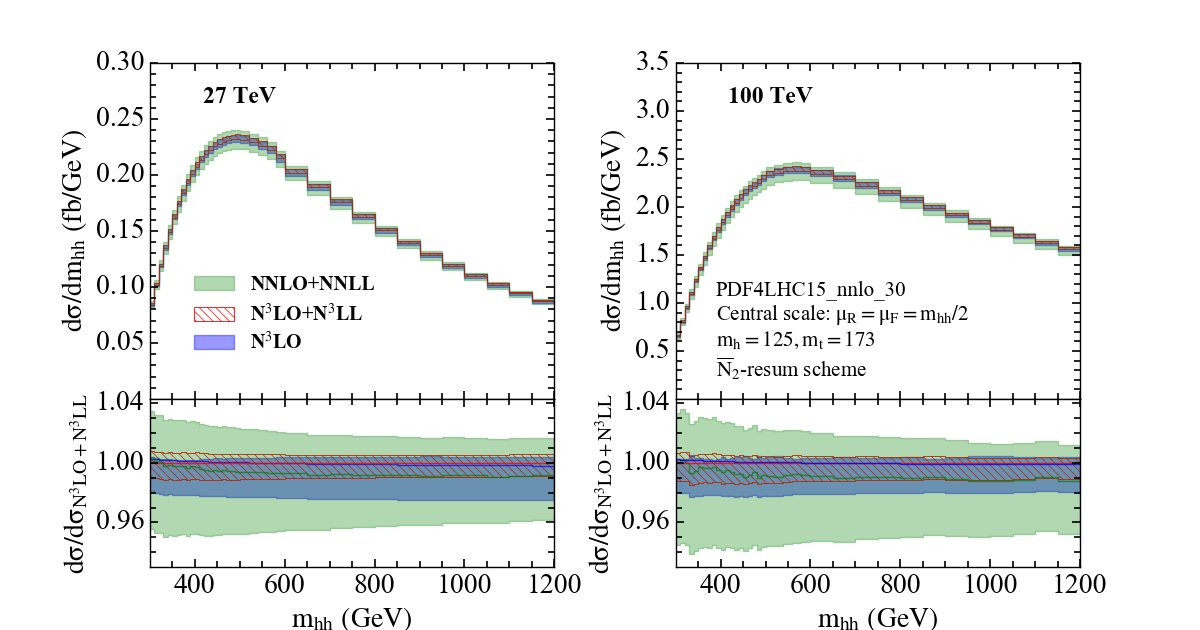}
\caption{The comparisons of the invariant mass differential cross sections $\frac{d\sigma}{dm_{hh}}$ between N$^3$LO+N$^3$LL (blue) and N$^3$LO (red hatched) and NNLO+NNLL (green) at four different center-of-mass energies $\sqrt{s}=13,14,27,100$ TeV. The error bands represent the 7-point scale variations with the central scale $\mu_0=\frac{m_{hh}}{2}$. The bottom panel shows the ratios to the N$^3$LO+N$^3$LL distribution.}
\label{fig:xs_vs_mhh_large_mt2}
\end{figure*}

In the infinite top mass limit, the only exactly known differential cross sections at N$^3$LO is the invariant mass $m_{hh}$ of the Higgs boson pair. While the other differential cross sections are computed only at some approximation at this order~\cite{Chen:2019fhs} so far. The invariant mass distribution is a typical observable for demonstrating the impact of threshold resummation calculations. In the $m_t\to\infty$ approximation, we report the $\frac{d\sigma}{dm_{hh}}$ distributions in figure~\ref{fig:xs_vs_mhh_large_mt1} at four distinct perturbative orders N$^k$LO+N$^k$LL ($k=0,1,2,3$). The inclusion of higher order QCD radiative corrections dramatically stabilises
the perturbative calculations of the invariant mass differential distributions. In particular, the shape of the distribution is almost unchanged from NLO+NLL to N$^3$LO+N$^3$LL except the $m_{hh}\simeq 2m_h$ regime, while the scale uncertainties are significantly reduced as what have been observed in the inclusive cross section case. The results in four nominal energies in the figure show that the $m_{hh}$ spectrum is harder from 13 TeV to 100 TeV because of the wider phase space with larger $\sqrt{s}$. From $k=1$, the error bands of N$^{k+1}$LO+N$^{k+1}$LL are completely enclosed within the previous-order N$^{k}$LO+N$^{k}$LL uncertainty bands. We compare our best N$^3$LO+N$^3$LL predictions to N$^3$LO and NNLO+NNLL in figure~\ref{fig:xs_vs_mhh_large_mt2}. The message is again similar. The N$^3$LL QCD corrections only marginally modify the shapes, and the N$^3$LO+N$^3$LL results with sub-percent
scale uncertainties are in both the N$^3$LO and NNLO+NNLL error bands.

\subsection{N$^3$LO+N$^3$LL improved cross sections with full top-quark mass dependence}

Unlike the single Higgs case, it is widely acknowledged that it would be crucial to include the finite top quark mass corrections in $gg\to hh$ for the realistic phenomenological applications. However, as aforementioned, the calculations in the full top quark mass dependence are very challenging, which are only best known at NLO+NLL~\cite{DeFlorian:2018eng}. In our paper, we will only use the NLO full $m_t$-dependent results (dubbed as ``$\NLOmt$" along the notation used in ref.~\cite{Chen:2019fhs}), which can be directly computed with the public code~\cite{Heinrich:2017kxx,Heinrich:2019bkc}. In order to provide the best theoretical predictions, a common way people usually do is to combine the $m_t\to\infty$ results, which are precisely known, with the full $m_t$ results. However, there are ambiguities on how to combine the two, featuring different underlying theoretical working assumptions. Several combination approaches have been studied in refs.~\cite{Chen:2019fhs,Maltoni:2014eza,Grazzini:2018bsd}. In this paper, we will opt to improving the (differential) cross sections by multiplying the (differential) QCD $K$ factors obtained in the $m_t\to \infty$ calculations, i.e.
\begin{eqnarray}\label{NkLO-fullmt}
d\sigma_{\NkLONkLLtimesNLOmt}&\equiv&d\sigma_{\NLOmt}\frac{d\sigma_{\NkLONkLL}}{d\sigma_{\rm NLO}},\nonumber\\
d\sigma_{\NkLOtimesNLOmt}&\equiv& d\sigma_{\NLOmt}\frac{d\sigma_{\NkLO}}{d\sigma_{\rm NLO}}.
\end{eqnarray}
Except the so-called ``finite top (FT)" approximation~\cite{Maltoni:2014eza,Grazzini:2018bsd}, which requires additional work that is beyond the scope of our paper, it is arguable that $\NkLONkLLtimesNLOmt$ (and $\NkLOtimesNLOmt$) is the best combination among those presented in section 3 of ref.~\cite{Chen:2019fhs}. Therefore, we refrain ourselves from presenting our results in various combination schemes here, but instead just show $\NkLONkLLtimesNLOmt$ and $\NkLOtimesNLOmt$. For $\NLOmt$ and $\NkLOtimesNLOmt$, the results are essentially identical to ref.~\cite{Chen:2019fhs} except that we use 7-point scale variations in the current paper, while 9-point scale variations have been adopted in ref.~\cite{Chen:2019fhs}. The cross section in each scale choice is evaluated following eq.(3.4) in ref.~\cite{Chen:2019fhs}. In other words, the relative scale errors in $\NkLONkLLtimesNLOmt$ and $\NkLOtimesNLOmt$ are identical to those in N$^k$LO+N$^k$LL and N$^k$LO in the infinite top quark mass approximation.  Though we have only taken into account $\NLOmt$, we want to emphasize that the most advanced theoretical predictions can be easily obtained from our calculation even after the full $m_t$ dependent NNLO result was available. In particular, we suggest the current recommendation values proposed by the LHC Higgs Cross Section Working Group~\cite{DiMicco:2019ngk} can be improved with our results.

The total cross sections after including $\NLOmt$ are listed in table~\ref{Tab:fullMt_inc} at the $4$ center-of-mass energies $\sqrt{s}=13,14,27$ and $100$ TeV. The results are obtained after applying eq.\eqref{NkLO-fullmt} to the total cross sections at NNLO+NNLL, N$^3$LO and N$^3$LO+N$^3$LL respectively. In the table, we show the original $\NLOmt$ results without any further improvement too. The residual scale uncertainties range from 10\% to 15\%. There is another source of big theoretical uncertainties due to the top-quark mass scheme and the scale arbitrariness in the $\overline{\rm MS}$ top quark mass that we have not yet quoted in the table, because such uncertainties are not expected to be reduced in our calculation with the infinite top quark mass approximation. In the inclusive total cross sections, these uncertainties are amount to $+4\%,-18\%$ at the above four energies~\cite{Baglio:2018lrj,Baglio:2020ini}. In the following, we will only focus on the conventional renormalisation and factorisation scale uncertainties. In the $\ResNbTwo$ resummation scheme, $\NNLONNLLtimesNLOmt$ enhances the central values of the $\NLOmt$ cross sections by around $20\%$, while the scale uncertainties have been reduced to $\sim \pm3\%$. Meanwhile, $\NtLOtimesNLOmt$ increases the $\NLOmt$ cross sections by a similar amount, but reduces the relative scale uncertainties to $-0.5\%,+2.5\%$, being very asymmetric and almost a factor $2$ smaller than $\NNLONNLLtimesNLOmt$. Our most advanced predictions $\NtLONtLLtimesNLOmt$ shrink such uncertainties by another factor of two, reaching to the sub-percent level. Our best predictions for the di-Higgs cross sections are $33.47, 39.60,151.9,1335$ fb at $\sqrt{s}=13,14,27,100$ TeV. We want to stress that though the central values of $\NNLONNLLtimesNLOmt$ are very close to $\NtLONtLLtimesNLOmt$, such a conclusion will alter if one chooses either $\ResNOne$ or $\ResNbOne$ resummation schemes.

\begin{table}[hbt!] 
\begin{center}
\begin{small}
\newcolumntype{P}[1]{>{\centering\arraybackslash}p{#1}}
{\renewcommand{\arraystretch}{1.5}
\begin{tabular}{|P{5cm} |P{2cm}|P{2cm}|P{2cm}|P{2cm}|}
\hline
     \multicolumn{1}{|c|}{ $\sqrt{s}$}   
    &\multicolumn{1}{c|}{ 13 TeV }  
    &\multicolumn{1}{c|}{14 TeV}  
    &\multicolumn{1}{c|}{27 TeV} 
    & \multicolumn{1}{c|}{100 TeV }\\
 \hline
 $\NLOmt$  &  27.56$_{-12.7\%}^{+13.9\%}$ & $32.64_{-12.47\%}^{+13.5\%}$& $126.1_{-10.4\%}^{+11.5\%}$& $1119_{-9.9\%}^{+10.7\%}$   \\
 \hline
 $\NNLONNLLtimesNLOmt$ &  $33.33_{-3.3\%}^{+3.0\%}$ & $39.42_{-3.4\%}^{+3.0\%}$ & $150.8_{-3.4\%}^{+2.7\%}$  & $1320_{-3.4\%}^{+2.4\%}$\\\hline
  $\NtLOtimesNLOmt$ & $33.43_{-2.8\%}^{+0.50\%}$&  $39.56_{-2.7\%}^{+0.50\%}$ &  $151.7_{-2.3\%}^{+0.46\%}$& $1333_{-1.8\%}^{+0.51\%}$ \\\hline
 $\NtLONtLLtimesNLOmt$ & $33.47_{-0.85\%}^{+0.88\%}$  & $39.60_{-0.87\%}^{+0.85\%}$  &$151.9_{-0.94\%}^{+0.63\%}$   & $1335_{-1.0\%}^{+0.35\%}$ \\
\hline
\end{tabular}}
\caption{The inclusive total cross sections (in unit of fb) of Higgs pair production at $\sqrt{s}=13,14,27,100$ TeV after taking into account the NLO full top quark mass dependence. The quoted relative uncertainties are from the 7-point scale variations.}
\label{Tab:fullMt_inc}
\end{small}
\end{center}
\end{table}

\begin{figure*}[htbp]
\centering
\includegraphics[width=1\textwidth]{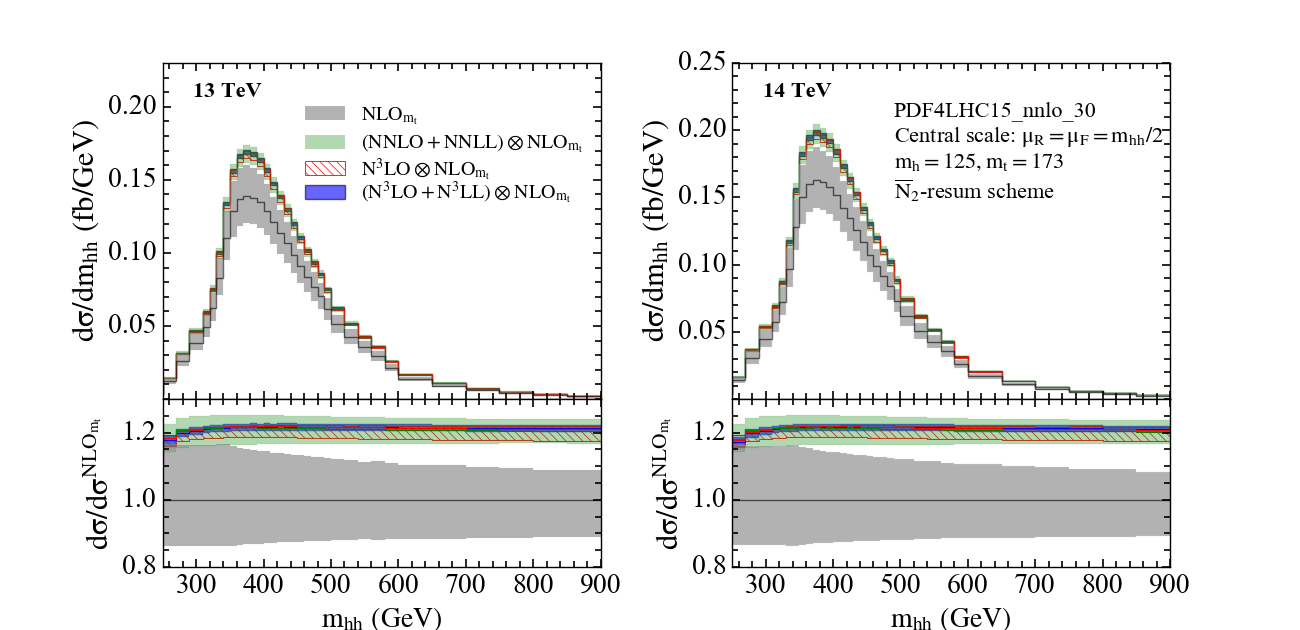}\\
\includegraphics[width=1\textwidth]{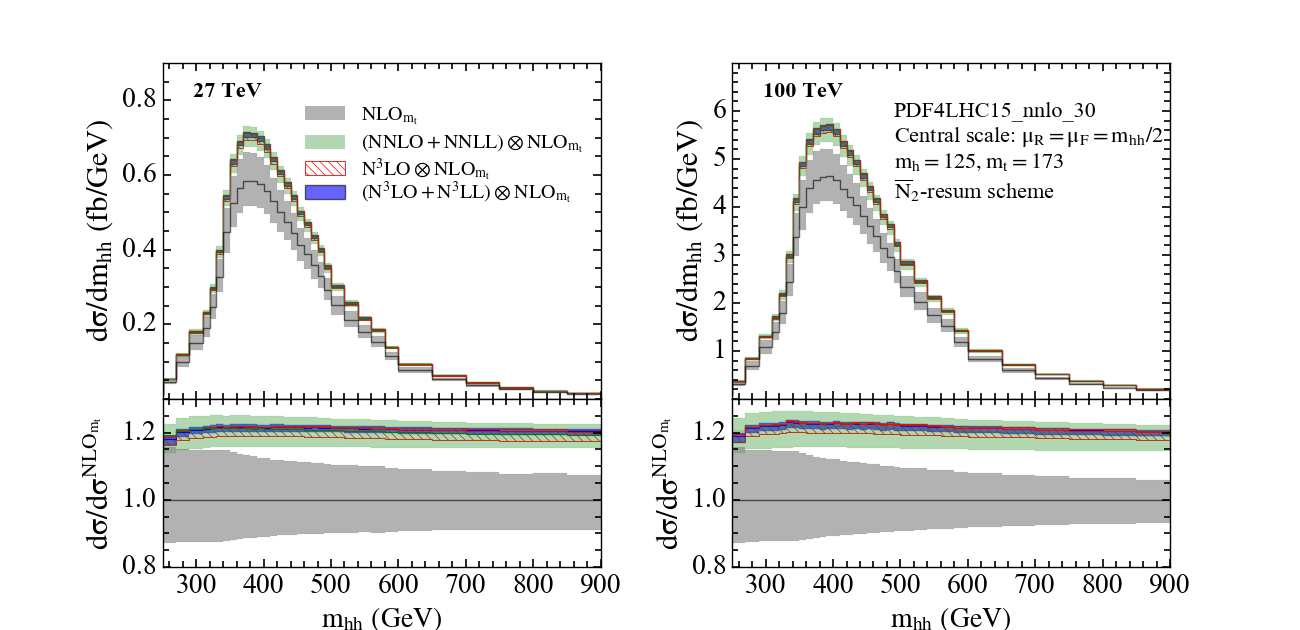}
\caption{The invariant mass $m_{hh}$ distributions with finite top quark mass corrections at four center-of-mass energies $\sqrt{s}=13,14,27,100$ TeV. The error bands are from the 7-point scale variations. The grey, green, red hatched and blue bands correspond to the $\NLOmt$, $\NNLONNLLtimesNLOmt$, $\NtLOtimesNLOmt$ and $\NtLONtLLtimesNLOmt$ predictions, respectively. The bottom panel shows the ratios to the $\NLOmt$ distribution.}
\label{fig:xs_vs_mhh_full_mt}
\end{figure*}

Figure~\ref{fig:xs_vs_mhh_full_mt} is our predictions for the invariant mass distributions after taking into account the NLO full top-quark mass dependence. Four theoretical bands in each plot are the $\NLOmt$ (grey bands), $\NNLONNLLtimesNLOmt$ (green bands), $\NtLOtimesNLOmt$ (red hatched bands), and $\NtLONtLLtimesNLOmt$ (blue bands) results. In principle, the infinite top quark mass approximation works better in the lower $m_{hh}$ regime. This is indeed  indicated in figure~\ref{fig:xs_vs_mhh_full_mt} that the $m_t\to\infty$ improved bands are only in the $\NLOmt$ bands when $m_{hh}<300$ GeV. Beyond $300$ GeV, the $K$ factors are almost constants around $1.2$, which can be evidently seen in the lower panels. Due to the manner of our combination, the remaining messages are essentially similar as what have been discussed in the previous subsection. 

Before closing this section, we assess the missing top quark mass uncertainties beyond NLO. This kind of uncertainty is stemming from the lack of full $m_t$-dependent calculations at NNLO and beyond. Since we are working in the infinite top quark mass approximation, we do not expect the inclusion of higher order in $\alpha_s$ will improve such an uncertainty. In other words, the missing top quark mass uncertainty at $\NtLONtLLtimesNLOmt$ should be as large as those at $\NNLOtimesNLOmt$ and $\NtLOtimesNLOmt$. The latter has been discussed in ref.~\cite{Chen:2019fhs} by comparing $\NNLOtimesNLOmt$ with respect to the so-called FT approximation at NNLO \cite{Grazzini:2018bsd}. This missing top quark mass uncertainty is found to be around 5$\%$ at the LHC energies to 9$\%$ at 100 TeV. Our best predictions at $\NtLONtLLtimesNLOmt$ is envisaged to be further improved by combining with the NNLO FT approximation calculations, which we leave it for the future work. 

\section{Summary\label{sec:conclusions}}

We have improved the (differential) cross section calculation of the Higgs boson pair production via gluon-gluon fusion with the threshold resummation up to N$^3$LL in the infinite top quark mass approximation. Such a resummation calculation has been consistently matched to the known N$^3$LO results. We first study the theoretical uncertainties at various perturbative orders due to the ambiguities in the resummation formalism. Though such resummation scheme uncertainties are in anyway sub-dominant at N$^3$LO+N$^3$LL, we argue that the $\ResNTwo$ and $\ResNbTwo$ schemes should give us the best predictions. With the $\ResNbTwo$ scheme, the residual renormalisation and factorisation scale uncertainties are less than one percent, which are a factor of $2$ smaller than N$^3$LO and a factor of $4$ smaller than NNLO+NNLL, while the central values are almost identical to N$^3$LO. This further consolidates the observation that the asymptotic convergence in the strong coupling constant $\alpha_s$ series has reached at the fourth order~\cite{Chen:2019fhs,Chen:2019lzz}. Such a conclusion holds for both the inclusive total cross sections and the differential invariant mass $m_{hh}$ distributions.

Our results have been finally combined with the full top quark mass dependent calculation at NLO QCD. Such combinations are essential for phenomenological applications. The new predictions of the di-Higgs hadroproduction cross sections have been reported after taking into account the finite top quark mass corrections.

Regarding the prospects for the future, it is clear that how to reduce the large top-quark mass scheme uncertainty remains an open question. In addition, the NNLO QCD and NLO electroweak calculations in the SM would be quite desirable, which are however quite technically challenging. The top-quark mass scheme uncertainty is expected to be reduced if the full NNLO QCD calculation in the SM was available. The minor percent level bottom-quark-loop contributions would turn out to be interesting too given the subpercent level scale uncertainty we have reached in this paper.

\section*{Acknowledgements}
We thank L.-B.~Chen, G.~Das, H.~T.~Li, P.~Mukherjee, V. Ravindran, and J. Wang for the collaboration at the initial stage of the project. We acknowledge the useful discussions with them at the different stages of this project. We are in particular grateful to H.~T.~Li for providing a grid of a two-loop amplitude and for checking the hard function against ref.~\cite{Chen:2019fhs}. This work is supported by the European Union’s Horizon 2020 research and innovation programme (grant agreement No.824093, STRONG-2020, EU Virtual Access``NLOAccess"), the ERC grant (grant agreement ID 101041109, ``BOSON"), the French ANR (grant ANR-20-CE31-0015, ``PrecisOnium"), and the CNRS IEA (grant No.205210, ``GlueGraph").

\newpage
\appendix
\section{The universal N$^3$LL resummed coefficients\label{app:A}}

In this appendix, we provide the explicit expressions of the process-independent ingredients in the resummed partonic coefficient function $\Delta_{gg}^{\rm res}(N,M^2,\mu_F^2)$ [eq.~\eqref{DeltaN}] for the gluon-gluon induced colourless final states in QCD with $n_q=5$. For the purpose of the discussions in subsection~\ref{sec:resscheme}, we would like to split $\tilde{C}_{0,gg}$ into two pieces
\begin{eqnarray}
    \tilde{C}_{0,gg}=\tilde{C}_{0,gg,\gamma_E}+\tilde{C}_{00,gg,\zeta},
\end{eqnarray}
where the first piece $\tilde{C}_{0,gg,\gamma_E}$ is proportional to the Euler-Mascheroni constant $\gamma_E$ and the second term solely depends on the Riemann zeta function which can be obtained by setting $\gamma_E=0$ in $\tilde{C}_{0,gg}$. The first $N$-independent universal part has the following $a_s$ series
\begin{align}
 2\ln {\cal S}_{gg,\delta}^{\rm fin} - 2\ln \Gamma_{gg,\delta}^{\rm fin} + \tilde C_{0,gg,\gamma_E}  = \sum_{k=1}{a_s^k~ G_{gg,0}^{(k)}}\,.
\end{align}
Up to three loops, we have
\begin{align}
    G_{gg,0}^{(1)} &= -   L_{fr}  \frac{46}{3} + 6 L_{qr}^2 -18 \zeta_2 + 24  \gamma_E (\gamma_E + L_{fr} - L_{qr} )\,,
    \\ 
    G_{gg,0}^{(2)} &=    L_{fr}^2  \frac{529}{9} 
  -   L_{fr} \Big(\frac{256}{3} +216  \zeta_3 \Big) 
  -  L_{qr}^3  \frac{46}{3} 
   +
   L_{qr}^2  \Big(\frac{302}{3}- 36 \zeta_2\Big) 
  -   L_{qr}  \Big(\frac{1864}{9} 
    \nonumber\\ &
  -   252 \zeta_3 - 184 \zeta_2\Big)
  +  \frac{5644}{27} 
  -
   \zeta_2 \frac{1057}{3}  +36 \zeta_2^2
   -   \zeta_3  \frac{782}{3} 
    + \gamma_E^3  \frac{368}{3}
    + \gamma_E^2  \bigg( \frac{1208}{3} - 184 L_{qr} - 144 \zeta_2 \bigg)
        \nonumber \\ &
 + \gamma_E  \bigg[ \frac{3728}{9} +    \Big(\frac{1208}{3}  - 144  \zeta_2\Big) ( L_{fr} -  L_{qr})
 - 92 (L_{fr}^2 - L_{qr}^2 )  - 504 \zeta_3 \bigg]\,,
   \\ 
    G_{gg,0}^{(3)} &= 
 L_{fr} \Big( 96  \zeta_2  - 276  \zeta_2^2-7248  \zeta_3 + 864 \zeta_2 \zeta_3+ 4320  \zeta_5  -  \frac{20881}{27} \Big)
 +  L_{fr}^2 \Big( 1656\zeta_3   +    \frac{2852}{3}\Big)
  \nonumber \\ &
  -
   L_{fr}^3  \frac{24334}{81} 
 +   L_{qr} \Big(    \zeta_2  \frac{78976}{9} - \zeta_2^2  \frac{8584}{5}+ \zeta_3  \frac{122768}{9} - 1584  \zeta_2 \zeta_3 - 5184 \zeta_5-   \frac{1472291}{243}  \Big)
 \nonumber\\&
 +  L_{qr}^2 \Big(  \frac{73097}{27} - \frac{7856}{3}  \zeta_2 +   \zeta_2^2 \frac{2376}{5}-2056\zeta_3  \Big)
  -   L_{qr}^3  \Big(\frac{15980}{27}-
 184  \zeta_2 \Big)
  +
   L_{qr}^4  \frac{529}{9} 
    \nonumber\\&
     +\frac{26820931}{4374}-
   \zeta_2  \frac{935587}{81}+ \zeta_2^2 \frac{107866}{45}
    +   \zeta_2^3  \frac{304}{7}  -   \zeta_3  \frac{1416988}{81} 
 +
   \zeta_2 \zeta_3  \frac{31724}{3} 
  \nonumber \\ &
  +
 3216 \zeta_3^2 -   \zeta_5 \frac{23816}{9} 
 +\gamma_E^4  \frac{8464}{9} + \gamma_E^3  \bigg( \frac{127840}{27} -  L_{qr}  \frac{16928}{9} - 1472 \zeta_2 \bigg) 
  \nonumber \\ &
+ \gamma_E^2  \bigg[ \frac{292388}{27} +  L_{qr} \Big(-\frac{63920}{9}  +
   2208 \zeta_2\Big) +   L_{qr}^2 \frac{4232}{3} - 4832 \zeta_2+   \zeta_2^2  \frac{9504}{5} - 8224 \zeta_3 \bigg]
   \nonumber\\ &
   +
 \gamma_E  \bigg[  L_{fr} \Big(\frac{40300}{9} -
   4832 \zeta_2 +  \zeta_2^2  \frac{9504}{5} - 496  \zeta_3\Big)
 + (L_{qr}^2 -L_{fr}^2 )\Big(\frac{31960}{9} - 1104  \zeta_2\Big)   
 \nonumber \\ &
 +   (L_{fr}^3 - L_{qr}^3 )  \frac{4232}{9} 
 +  L_{qr} \Big(-\frac{292388}{27} + 4832  \zeta_2-   \zeta_2^2  \frac{9504}{5}+ 8224  \zeta_3\Big)
 +  \frac{2944582}{243} 
\nonumber\\&
 -   \zeta_2 \frac{30112}{9}   -
     \zeta_2^2  \frac{4912}{5} -
     \zeta_3  \frac{177824}{9}  + 3168 \zeta_2 \zeta_3 +
   10368 \zeta_5 \bigg]\,,
\end{align}
where we have defined $L_{qr}\equiv\ln \frac{M^2}{\mu_R^2}$ and $L_{fr}\equiv\ln \frac{\mu_F^2}{\mu_R^2}$. In the above equations, the $\gamma_E$ terms are purely from $\tilde{C}_{0,gg,\gamma_E}$. Therefore, the expressions for $2\ln {\cal S}_{gg,\delta}^{\rm fin} - 2\ln \Gamma_{gg,\delta}^{\rm fin}$ can be obtained directly by setting $\gamma_E=0$. For $\tilde{C}_{0,gg,\zeta}$, we have the similar $a_s$ series $ \tilde C_{0,gg,\zeta} = \sum_{k=1} a_s^k \tilde{C}_{0,gg,\zeta}^{(k)}$ with the coefficients
\begin{eqnarray}
    \tilde{C}_{0,gg,\zeta}^{(1)} &&= 24 \zeta_2\,,
\\
 \tilde{C}_{0,gg,\zeta}^{(2)}&& =     \zeta_2  \frac{1208}{3} -
 184 L_{qr} \zeta_2 - 144 \zeta_2^2   +
   \zeta_3  \frac{736}{3}\,,
\\ 
  \tilde{C}_{0,gg,\zeta}^{(3)}&& =   
     L_{qr} \Big(-\zeta_2  \frac{63920}{9}+ 2208  \zeta_2^2-
   \zeta_3  \frac{33856}{9}\Big) +  L_{qr}^2 \zeta_2  \frac{4232}{3} +
   \zeta_2  \frac{292388}{27}-
   \zeta_2^2  \frac{80944}{15}  
   \nonumber \\& &
   +   \zeta_2^3  \frac{9504}{5} 
   +   \zeta_3  \frac{255680}{27}  - 11168 \zeta_2 \zeta_3\,.
\end{eqnarray}

The $N$-dependent coefficients as given in the exponent of eq.~\eqref{DeltaN} are
\begin{align}
    g_{1,gg}(\omega) &= \frac{72}{23}\frac{1}{\omega} \Big[(1-\omega)\ln(1-\omega) + \omega\Big]\,,
\\
    g_{2,gg}(\omega) &= 
\ln^2(1-\omega) \frac{  6264 }{12167} 
 +
\ln(1-\omega)  \bigg( \frac{ -29148}{12167} - \frac{  72 \gamma_E  }{23} + \frac{  36 L_{qr}  }{23} + \frac{ \ 648 \zeta_2  }{529} \bigg)
\nonumber \\ &
+ \omega  \bigg( \frac{-29148}{12167} + \frac{  36 L_{fr}  }{23} + \frac{  648 \zeta_2  }{529} \bigg) 
\,,
\\
    g_{3,gg}(\omega) &= \frac{1}{ (-1+\omega) }   \bigg[\frac{ -726624  }{279841}\ln^2(1-\omega) + \ln(1-\omega)  \bigg(\frac{ 151578}{12167} + \frac{  8352 \gamma_E  }{529} -
   \frac{  4176 L_{qr}  }{529} - \frac{ 75168 \zeta_2  }{12167} \bigg)  
   \nonumber \\ &
+\omega^2  \bigg( \frac{ -2965030}{279841} + \frac{  604 L_{fr} }{23} - 6 L_{fr}^2 + \frac{  212472 \zeta_2  }{12167} -
   \frac{  216 L_{fr} \zeta_2 }{23} - \frac{ 21384 \zeta_2^2 }{2645} + \frac{  1116 \zeta_3  }{529 }\bigg)  
   \nonumber \\ &
 +\omega  \bigg(\frac{ -531322}{36501} - \frac{ 19432 \gamma_E  }{529} - 24 \gamma_E^2 - \frac{  604 L_{fr}  }{23} + 6 L_{fr}^2 +
   \frac{ 9716 L_{qr} }{529} + 24 \gamma_E L_{qr} 
   - 6 L_{qr}^2 
    \nonumber \\ &
   - \frac{ 105126 }{279841} \ln(1-\omega)  -
   \frac{  75168 \zeta_2 }{12167} + \frac{  432 \gamma_E \zeta_2  }{23} + \frac{ 216 L_{fr} \zeta_2  }{23} -
   \frac{ 216 L_{qr} \zeta_2  }{23} + \frac{  756 \zeta_3  }{23} \bigg)  \bigg]\,, 
 \end{align}
  \begin{align}
   g_{4,gg}(\omega) &= \frac{1}{ \big(-1+\omega \big)^2} \bigg\{ \frac{ -28096128 }{6436343}  \ln^3(1-\omega) + \ln^2(1-\omega)  \bigg(\frac{ 12191136}{279841} + \frac{  484416 \gamma_E  }{12167} 
   \nonumber \\ &
   -\frac{  242208 L_{qr}  }{12167} 
   - \frac{  4359744 \zeta_2  }{279841} \bigg)  +
 \ln(1-\omega)  \bigg(\frac{ -519707665}{2518569} - \frac{  140128 \gamma_E  }{529} - \frac{  2784 \gamma_E^2  }{23} 
    \nonumber \\ &
 +
   \frac{  70064 L_{qr}  }{529} 
   + \frac{  2784 \gamma_E L_{qr}  }{23} - \frac{  696 L_{qr}^2  }{23} +
   \frac{  24646752 \zeta_2 }{279841} + \frac{ 50112 \gamma_E \zeta_2  }{529} - \frac{  25056 L_{qr} \zeta_2  }{529}
      \nonumber \\ &
   -
   \frac{  2480544 \zeta_2^2  }{60835} 
   + \frac{  2852192 \zeta_3  }{12167} \bigg)  +
 \omega^2  \bigg[\frac{-3406003448}{22667121} - \frac{ 63148756 \gamma_E  }{109503} - \frac{  1208 \gamma_E^2  }{3} 
  \nonumber \\ &
 -
   \frac{ 368 \gamma_E^3  }{3} 
   - \frac{  40300 L_{fr}  }{69} + \frac{  31960 L_{fr}^2  }{69} - \frac{  184 L_{fr}^3}{3} +
   \frac{ 31574378 L_{qr}  }{109503} + \frac{ 1208 \gamma_E L_{qr}  }{3} 
     \nonumber \\ &
   + 184 \gamma_E^2 L_{qr} 
  - \frac{ 302 L_{qr}^2  }{3} -
   92 \gamma_E L_{qr}^2 + \frac{ 46 L_{qr}^3 }{3} - \frac{  203879104 \zeta_2 }{839523} +
   \frac{  141648 \gamma_E \zeta_2  }{529} 
    \nonumber \\ &
   + 144 \gamma_E^2 \zeta_2
   + \frac{ 14496 L_{fr} \zeta_2  }{23} -
   144 L_{fr}^2 \zeta_2 - \frac{  70824 L_{qr} \zeta_2  }{529} - 144 \gamma_E L_{qr} \zeta_2 +
   36 L_{qr}^2 \zeta_2
      \nonumber \\ &
   + \frac{ 26054472 \zeta_2^2 }{60835}
   - \frac{  14256 \gamma_E \zeta_2^2  }{115} -
   \frac{  28512 L_{fr} \zeta_2^2  }{115} + \frac{  7128 L_{qr} \zeta_2^2  }{115} -
   \frac{  3036528 \zeta_2^3  }{18515}  
   \nonumber \\ &
   + \frac{  14431616 \zeta_3  }{36501}
   + \frac{  12336 \gamma_E \zeta_3  }{23} +
   \frac{ 1488 L_{fr} \zeta_3 }{23} - \frac{ 6168 L_{qr} \zeta_3  }{23} - \frac{  41256 \zeta_2 \zeta_3  }{529} -
   \frac{ 34992 \zeta_3^2  }{529} 
   \nonumber \\ &
   + \ln(1-\omega)  \bigg( \frac{ -965392751}{57927087} + \frac{  705728 \zeta_3 }{12167 } \bigg) -
   \frac{  98592 \zeta_5 }{529  }\bigg] 
   + \omega^3  \bigg(\frac{ -5557403702}{57927087} + \frac{  20150 L_{fr}  }{69} 
   \nonumber \\ &
   -
   \frac{  15980 L_{fr}^2 }{69} + \frac{  92 L_{fr}^3  }{3} + \frac{  172965524 \zeta_2  }{839523} -
   \frac{ 7248 L_{fr} \zeta_2 }{23} + 72 L_{fr}^2 \zeta_2 - \frac{  15243576 \zeta_2^2 }{60835} +
   \nonumber \\ &
   \frac{  14256 L_{fr} \zeta_2^2  }{115} + \frac{  2024352 \zeta_2^3  }{18515} +
   \frac{  3204880 \zeta_3  }{36501} - \frac{ 744 L_{fr} \zeta_3 }{23} - \frac{  8928 \zeta_2 \zeta_3 }{529} +
   \frac{  23328 \zeta_3^2  }{529} 
   \nonumber \\ &
   - \frac{  53504 \zeta_5  }{529 }\bigg)  +
 \omega  \bigg[\frac{ 13235995612}{22667121} + \frac{ 5463772 \gamma_E  }{4761} + \frac{  2416 \gamma_E^2  }{3} + \frac{  736 \gamma_E^3  }{3} +
   \frac{  20150 L_{fr}  }{69} 
   \nonumber \\ &
   - \frac{ 15980 L_{fr}^2 }{69} + \frac{  92 L_{fr}^3  }{3} - \frac{  2731886 L_{qr} }{4761} -
   \frac{  2416 \gamma_E L_{qr}  }{3} - 368 \gamma_E^2 L_{qr} + \frac{  604 L_{qr}^2  }{3} 
   \nonumber \\ &
   + 184 \gamma_E L_{qr}^2 -
   \frac{ 92 L_{qr}^3 }{3} - \frac{ 109246096 \zeta_2  }{839523} - \frac{  283296 \gamma_E \zeta_2  }{529} -
   288 \gamma_E^2 \zeta_2 - \frac{ 7248 L_{fr} \zeta_2  }{23} 
  \nonumber \\ &
   + 72 L_{fr}^2 \zeta_2 +
   \frac{  141648 L_{qr} \zeta_2  }{529} + 288 \gamma_E L_{qr} \zeta_2 - 72 L_{qr}^2 \zeta_2 -
   \frac{  6378216 \zeta_2^2  }{60835} + \frac{28512 \gamma_E \zeta_2^2  }{115} 
   \nonumber \\ &
   +
   \frac{ 14256 L_{fr} \zeta_2^2  }{115} - \frac{  14256 L_{qr} \zeta_2^2  }{115} +
   \ln(1-\omega)  \bigg(\frac{ 88719002}{2518569} - \frac{  1411456 \zeta_3  }{12167 } \bigg)
   - \frac{  38477872 \zeta_3 }{36501}
   \nonumber \\ &
   -
   \frac{ 24672 \gamma_E \zeta_3  }{23} - \frac{ 744 L_{fr} \zeta_3  }{23} + \frac{ 12336 L_{qr} \zeta_3  }{23} +
   \frac{ 4752 \zeta_2 \zeta_3  }{23} + \frac{  15552 \zeta_5  }{23  }\bigg]\bigg\}\,.
\end{align}
\bibliographystyle{JHEP}
\bibliography{bib}
\end{document}